_______________________________________________________________________________
\documentstyle[preprint,eqsecnum,aps]{revtex}
\begin{document}
\draft
\title{
VECTOR COLOR TRANSPARENCY}
\author{W.~R.~Greenberg\cite{present} and G.~A.~Miller}
\address{
Department of Physics, FM-15\\
 University of Washington, Seattle, WA 98195}
\maketitle
\begin{abstract}
Color transparency (CT) in high momentum transfer $(e,e' \vec p)$
reactions is explored. The spin of the proton and photon are treated
explicitly, hence the name ``Vector CT".
The Dirac distorted wave impulse approximation is used as a
starting point; then CT effects are embedded.
A hadronic basis is used to describe the
struck proton as a wavepacket of physical baryon resonances.
The effects of the wavepacket expansion on the normal component of the
ejectile polarization, which vanishes in the limit of full CT, are
investigated.
This formalism is also applied to study CT effects in total cross sections,
individual separated nuclear response functions,
Fermi motion of the initial nucleon, non-zero size of the initial wavepacket
and the
effects of relativistic lower components.  We show that including
CT reduces the violations of current conservation (CC), a typical problem
in calculations of this kind.  The energy dependence of the normal
polarization in $(e,e' \vec p)$ reactions is found to be slow.
However, a
measurement of the normal transverse response in a heavy nucleus, such as
$^{208}Pb$ seems to afford the opportunity to see CT at quite low
momentum transfers.  The effects of Fermi motion are investigated, and
choosing the momentum of the struck nucleon to be large leads to
significant violations of CC.
\end{abstract}

\narrowtext
\section{INTRODUCTION}
\label{sec:intro}

Color Transparency (CT) is the postulated~\cite{MU82,BR82}
absence of final- (or initial-) state interactions
caused by the cancellation of
color fields of a system of quarks and gluons
with small spatial separation.
For example, suppose an electron impinges on a
nucleus knocking out a
proton at high momentum transfer.  The consequence of
color transparency is that
there is no exponential loss of flux as the ejected
particle propagates through the
nucleus.  Thus, the usually ``black" nucleus
becomes transparent. We restrict our attention to
processes for which the fundamental reaction is elastic, or at least a
two-body reaction.  This requires that the  nuclear excitation energy
be known well enough
to ensure that no extra pions are created.
This subject is under active experimental
investigation~\cite{CA88,HE89,BNL,NE18,McK92,CEBAF,EEF,EFGMSS}.

  The existence of color transparency depends on: (1)~the formation of a
small-sized wave-packet in a high momentum transfer reaction. (2)~the
interaction between such a
small object and nucleons being suppressed (color
neutrality or screening)
and (3)~the wave-packet escaping the nucleus while
still small~\cite{MU82,BR82}.
That color neutrality (screening)
causes the cross section of small-sized color singlet
configurations with hadrons to be small was found in
Refs.~\cite{LO75,NU75,GS77,PH87,BBGG81}, and is
well-reviewed in Refs.~\cite{FS88,KO90,FS91,FMS92}.
So we take item (2) as given. The truth of item (1), for experimentally
available energies,
is an interesting issue. It is discussed in
Refs.~\cite{FMS92,LS92,FMS92c}, and
is not probed in depth here.

It is also true that, for present
experiments~\cite{CA88,HE89,BNL,NE18,McK92,CEBAF,EEF,EFGMSS},
the small object does expand as it moves
through the nucleus.
Thus the final-state interactions are suppressed but not
zero. The importance of this expansion was found by Farrar et
al.~\cite{FLFS88}, and by
Jennings and Miller~\cite{JM90,JM91,MJ92,MI92}. See also
Refs.~\cite{KZ91a,KZ91b}.

Up until now, all calculations of CT have
 totally neglected the effects of spin.
In this paper, the full Lorentz structure of the
matrix element for the electroproduction of nucleons from nuclei is
considered.
We call this work
Vector Color Transparency,
where the vector moniker derives from the vector
nature of the photon
exchanged in the hard collision.

What do we hope to gain by introducing this complication?  First,
experience has shown that the introduction of spin into many physical
problems led to totally new phenomena.  Whether or not this happens in
discussions of CT is the subject of this paper.  Second, by
invoking time-reversal and parity invariance of the electromagnetic and
strong interactions, the normal component of the
polarization of the ejected proton [ in $(e, e' {\vec p})$ experiments]
must vanish in the
limit where final-state interactions are absent~\cite{Zhalov}. This
is the limit of full CT.
Therefore, the spin dependent observables provide
a very sensitive measure of the effects of CT.  Thirdly,
we will soon see that the
spin dependent observables are defined in terms of
ratios of sums of response
functions.  Thus, experimental measurement of such
observables are less susceptible to
systematic errors than, for instance,
the unpolarized observables.
Lastly, an experiment has been proposed and
approved, to be run at CEBAF~\cite{CEBAF},
which will measure the energy dependence of
the unpolarized
cross section and the
normal component of the ejected proton's
polarization.  In that case, it is desirable to have ready theoretical
predictions, crude though they might be, to be confronted by the
experimental results.

In a previous paper~\cite{GM93}
a multiple-scattering series
for the color transparency wavepacket-nucleon interaction, within the
framework of Glauber (eikonal) theory, was developed.
The approach here is the
same, except that four-component Dirac spinors are used for both scattered
wavefunctions and bound states.
As in Ref.~\cite{GM93}, we embed CT operators and states
into the usual DWBA treatment.

The organization of this paper is as follows.  In Section II
we display formulae for
the $(e,e' \vec p)$ cross section and polarization
in terms of the nuclear response
functions, the hadronic tensor, and the
nuclear current matrix element (NCME).  In Section III
the scattering state wavefunction is derived within the context of the
Dirac impulse approximation.  Most of the material in Sections II and III
is well-known, but we present these sections as necessary preliminaries
to including the effects of CT.  This is done in Section~\ref{ch:incvct},
by including the effects of the composite nature of the nucleon.
We develop approximation schemes,
as in Ref.~\cite{GM93}, to evaluate the
CT wavefunctions, in Section~\ref{ch:eval2}.  Section VI contains a
demonstration that there is at least one reliable
approximation
scheme for each range of energies.  In this section, we
also present numerical results for
total cross section ratios,
integrated longitudinal and transverse nuclear
responses, current conservation violations,
differential unpolarized cross sections,
differential normal polarizations,
individual nuclear response functions,
effects of non-zero wavepacket
size, Fermi motion, and the effects of the purely relativistic lower
components of the bound and scattered state wavefunctions.
In the final section, Section VII,
we summarize and make
some concluding comments.

The publication of the results of the SLAC experiment~\cite{NE18}  is imminent
but
not published now.  It is important to use the experimental acceptance
in computing observables.
Thus we leave a detailed assessment of that
experiment to a future publication.
We also make no attempt to completely
review the CT literature.

\section{The Nuclear Current Matrix Element and Cross Section}
\label{ch:vncme}
We now describe the spin-dependent formalism.
In particular, consider the
case of bombarding a
spin-0 nucleus with polarized electrons and
detecting polarized protons
which are knocked out.
This process is usually written as $(\vec e, e' \vec p)$.  The
incoming electron has helicity $h$ and the
outgoing nucleon is polarized, in its rest frame,
along $\check{\bf s}_R$.  A schematic drawing of the
$(\vec e, e' \vec p)$ reaction is shown in Figure~\ref{fig:veep}.
The following discussion reproduces the essential
points of Ref.~\cite{PVO87}

The kinematical situation is
shown in Figure~\ref{fig:kinematics}.
We take the virtual photon to lie along the $\check{\bf Z}$
direction, and the electron scattering plane to be the YZ plane.
The scattering amplitude for Vector CT is defined as
\begin{equation}
M_\alpha = j_\mu J^\mu_\alpha, \label{eq:mjjv}
\end{equation}
where $j_\mu$ is the matrix element of the electron
current and $J^\mu_\alpha$ is
the matrix element of the nuclear current (NCME).
In the
one-photon exchange approximation and using the nuclear shell model,
the cross section is
\begin{eqnarray}
& \frac{d^3 \sigma}{d \epsilon_{k'} d \Omega_{k'} d\Omega_p} &=
\frac{M_N |{\bf p}|}{(2\pi)^3}
\left( \frac{d\sigma}{d\Omega_{k'}} \right)_{Mott}\nonumber\\
&\times &
\sum_\alpha \int {dE_p}
| M_\alpha |^2 \delta(E_p - q^0 -M_N + \varepsilon_\alpha),
\label{eq:crossv}
\end{eqnarray}
where $k^\mu ({k'}^\mu)$ is the
four-momentum of the incoming (outgoing)
electron, $q^\mu=k^\mu-{k'}^\mu$.
The binding energy of
the nucleon in shell $\alpha$ is $\varepsilon_\alpha$.  This is small
compared with other energy scales in the problem and is
neglected.  The sum is over all occupied shells.
The four-momentum of the outgoing nucleon is $p^\mu$ and its
energy is denoted by $p^0=E_p$; $M_N$ is the nucleon
mass.  The solid angle $d\Omega_p = \sin \zeta \, d\zeta \, d\beta$.
The nucleus is assumed to
recoil with negligible energy.
The Mott cross section is defined in Ref.~\cite{PVO87}

The square of the matrix element $M_\alpha$ can be written in terms of
electron and nuclear tensors:
\begin{equation}
|M_\alpha|^2 = \eta_{\mu \nu} W^{\mu \nu}_\alpha,
\label{eq:etaw}
\end{equation}
The electron tensor
$\eta_{\mu \nu} = \frac{1}{2}
\left( k_\mu {k'}_\nu + k_\nu {k'}_\mu
- g_{\mu \nu} k \cdot k' +
ih \epsilon_{\mu \nu \lambda \kappa}{k'}^\lambda
k^{\kappa} \right)$,
where $g_{\mu \nu}$ is the metric tensor with $g_{00}=-g_{11}=-g_{22}=-
g_{33}=1$, and $\epsilon_{\mu \nu \alpha \beta}$ is
the completely antisymmetric fourth rank tensor.  The electron helicity
is given by  $h$.  Here and throughout we use the notation and conventions
of Bjorken and Drell~\cite{BD}.

We turn now to the nuclear tensor.
This depends on the direction of
the (rest-frame) spin of the ejected proton.  Thus,
$W^{\mu \nu}_\alpha
= W^{\mu \nu}_\alpha(\check{\bf s}_R)$
where we use the notation that
$\check{\bf a}$ describes a unit vector in the direction of $\bf a$.
The upside-down ``hat'' is used to distinguish unit vectors from operators.
The nuclear tensor is
\begin{equation}
W^{\mu \nu}_\alpha
(\check{\bf s}_R) =
J^{\mu *}_\alpha (q) J^\nu_\alpha (q),
\label{eq:wmunu}
\end{equation}
where
$J^\nu_\alpha (q)$
are the matrix elements
of the nuclear electromagnetic current operator.
All of the unknown physics is lumped
into the nuclear current.  Therefore, we define the
quantity,
\begin{equation}
\overline{W}^{\mu \nu}
\left( {\check{\bf s}_R} \right) = \sum_\alpha \int dE_p \,
W^{\mu \nu}_\alpha
\left( {\check{\bf s}_R} \right) \delta \left( E_p - q^0 -M_N +
\varepsilon_\alpha \right),
\label{eq:wbar}
\end{equation}
where the sum is over all
occupied single-particle shells and the integral
and delta function serve to conserve energy in this full nuclear tensor.
The nuclear tensor $W^{\mu \nu}_\alpha$ depends on the ejected proton
energy, $E_p$, because the scattered proton is ultimately detected, on
the mass shell, moving with momentum $\bf p$.

Thus, the crucial
quantity is the nuclear current matrix element (NCME).
We now present a
symbolic form for this current.  The NCME is given by
\begin{equation}
J^\mu_\alpha(q)
=  \, ^{(-)} \! \langle \Psi_{{\bf p}, \check{\bf s}_R} | T_H^\mu(q)
| \alpha \rangle.
\label{eq:jmu}
\end{equation}
where the initial state of this knockout
process is labelled by the shell model state of a bound proton, $\alpha$.
In the final state, a proton moves with a momentum
$\bf p$ with some rest frame spin $\check{\bf s}_R$.  The residual nucleus
is an $\alpha$-hole state, which is not detected and is therefore summed
over.
The initial and final states are connected by
a vector operator denoted as $T^\mu_H(q)$ which describes the absorption
of a virtual photon on a
proton in the nucleus.
The overlap of the initial and final nuclear states is imagined to be a
single particle state of a nucleon bound in shell model state $\alpha$.
Incoming boundary
conditions for the proton scattered wave are used, but
we can also use
outgoing boundary conditions:
\begin{equation}
J^\mu_\alpha(q)
= \langle \alpha | T_H^{\mu \dagger}(q) | \Psi_{{\bf p}, -\check{\bf s}_R}
\rangle^{(+)}.
\label{eq:jmuout}
\end{equation}
A more complete
derivation is described elsewhere~\cite{PVO87}.

We contract the indices of the electron and nuclear tensors to obtain the
differential cross section.  The use of current conservation:
$q_\mu J^\mu =
q^0 J^0 - q^3 J^3 = 0$, where ${\bf q} \equiv q^3 \check{\bf Z}$
allows us to eliminate the longitudinal
components of the nuclear current in favor of the better known
charge density.  In
particular,
$W^{33}=\frac{q_0^2}{q_3^2}W^{00}$ and $W^{3k}=\frac{q_0}{q_3}
W^{0k}$ for $k \neq 3$.  We note here an important point concerning current
conservation and gauge invariance.  The NCME is an integral over a bound
state wavefunction, a scattered state wavefunction and an operator which
connects the two.  If these two wavefunctions are derived from the same
theory, as they are in nature, then as members of a complete set they are
orthogonal and current is conserved.  If the bound and
scattered states are modelled with different Hamiltonia, as usually
practiced, there is no guarantee that the wavefunctions are
orthogonal and that current is conserved.
This problem is discussed in Section~\ref{ch:eval2} where we show
that current is conserved in our
calculation at the $<10$\% level or better.

The above discussion leads
to the differential cross section for the scattering of polarized
electrons (helicity $h$) off of a nucleus, ejecting a proton with rest-
frame spin $\check{\bf s}_R$~\cite{PVO87}:
\widetext
\begin{eqnarray}
\sigma_h(\check{\bf s}_R) & \equiv &
\left( { {d^3\sigma}\over{d\epsilon_{k'}
d\Omega_{k'} d\Omega_{p} }}\right)_{h,\check{\bf s}_R}\\
 &=& \frac{M_N |{\bf p}|}{2 (2\pi)^3}
\left( { {d\sigma}\over{d\Omega_{k'}}}\right)_{Mott}  \Bigl\{
V_L \left( R_L + R_L^n {\cal S}_n \right)
 + V_T \left( R_T + R_T^n {\cal S}_n \right) \nonumber \\*
&& + V_{TT} \bigl[ \left(R_{TT}+R_{TT}^n {\cal S}_n \right) \cos 2\beta +
\left(R_{TT}^t {\cal S}_t + R_{TT}^l {\cal S}_l \right) \sin 2\beta \bigr]
\nonumber \\*
&& + V_{LT} \bigl[ \left(R_{LT}+R_{LT}^n {\cal S}_n \right) \sin \beta +
\left(R_{LT}^t {\cal S}_t + R_{LT}^l {\cal S}_l \right) \cos \beta \bigr]
\nonumber \\*
&& + h V_{LT}' \bigl[ \left(R_{LT}'+R_{LT}^{\prime n} {\cal S}_n \right)
\cos \beta +
\left(R_{LT}^{\prime t} {\cal S}_t + R_{LT}^{\prime l} {\cal S}_l \right)
\sin \beta \bigr]
\nonumber \\*
&& + h V_{TT}'
\left( R_{TT}^{\prime t} {\cal S}_t + R_{TT}^{\prime l} {\cal S}_l \right)
\Bigr\},
\label{eq:eepcs}
\end{eqnarray}
\narrowtext
where kinematical factors $V$ are defined by
$V_L = \left( \frac{q^2}{|{\bf q}|^2} \right)^2$,
$V_T = \tan^2 \frac{\theta_e}{2} - \frac{q^2}{2|{\bf q}|^2}$,
$V_{TT} = - \frac{q^2}{2|{\bf q}|^2}$,
$V_{LT} = \frac{q^2}{|{\bf q}|^2}
\left(\tan^2 \frac{\theta_e}{2} - \frac{q^2}{|{\bf q}|^2}\right)^{1/2}$,
$V_{LT}' = \frac{q^2}{|{\bf q}|^2} \tan \frac{\theta_e}{2}$,
$V_{TT}' = \tan \frac{\theta_e}{2}
\left(\tan^2 \frac{\theta_e}{2} - \frac{q^2}{|{\bf q}|^2}\right)^{1/2}$,
where $\theta_e$ is the electron scattering angle, $q^2 = q_\mu q^\mu$, and
$|{\bf q}|$ is the magnitude of the photon three-momentum.  The direction of
the ejected proton's spin is given, in its rest frame, by $\check{\bf s}_R$.
The spin projections which appear in the above cross section are identified
to be:
${\cal S}_n  =  \check{\bf n} \cdot \check{\bf s}_R$,
${\cal S}_l  =  \check{\bf l} \cdot \check{\bf s}_R$,
${\cal S}_t  =  \check{\bf t} \cdot \check{\bf s}_R$,
where the unit vectors are
$\check{\bf n}=
\check{\mbox{\boldmath $\beta$}} = \left( {\bf q} \times {\bf p} \right) /
|{\bf q} \times {\bf p}|$,
$\check{\bf l}=\check{\bf p}$
and $\check{\bf t}=
\check{\mbox{\boldmath $\zeta$}} = \check{\bf n} \times \check{\bf l}$ .
The quantities $R$ are called the response functions.  The subscripts $L$
and $T$ are the contributions from longitudinal and transverse photons
(both helicities),
respectively.  The subscripts $TT$ and $LT$ refer to the transverse-
transverse and longitudinal-transverse response of the system.  These arise
from the interference between the different polarizations of the photon.
The subscripts $TT'$ and $LT'$ also arise from the
interference of the longitudinal and transverse photons.  They are
``primed'' because they are formed from an antisymmetric combination of
components of the nuclear tensor and thus only accessible with a polarized
electron beam.
The response functions themselves are defined in terms of the
hadronic tensor.  Explicitly,
\widetext
\begin{equation}
\frac{1}{2} \left( R_L + R_L^n {\cal S}_n \right) =
\overline{W}^{00}(\check{\bf s}_R),
\label{eq:rlrln}
\end{equation}
\begin{equation}
\frac{1}{2} \left( R_T + R_T^n {\cal S}_n \right) =
\overline{W}^{11}(\check{\bf s}_R)+\overline{W}^{22}(\check{\bf s}_R) ,
\end{equation}
\begin{equation}
 \frac{1}{2}  \bigl[
\left(R_{TT}+R_{TT}^n {\cal S}_n \right) \cos 2\beta
 +
\left(R_{TT}^t {\cal S}_t + R_{TT}^l {\cal S}_l \right) \sin 2\beta
\bigr] =
\overline{W}^{22}(\check{\bf s}_R)-\overline{W}^{11}(\check{\bf s}_R) ,
\end{equation}
\begin{equation}
 \frac{1}{2}  \bigl[
\left(R_{LT}+R_{LT}^n {\cal S}_n \right) \sin\beta
 +
\left(R_{LT}^t {\cal S}_t + R_{LT}^l {\cal S}_l \right) \cos\beta
\bigr]  =
\overline{W}^{02}(\check{\bf s}_R)+\overline{W}^{20}(\check{\bf s}_R) ,
\end{equation}
\begin{equation}
 \frac{1}{2}  \bigl[
\left(R_{LT}'+R_{LT}^{\prime n} {\cal S}_n \right) \cos \beta
+
\left(R_{LT}^{\prime t} {\cal S}_t + R_{LT}^{\prime l} {\cal S}_l \right)
\sin \beta \bigr]  =
i \left[\overline{W}^{10}(\check{\bf s}_R)-
\overline{W}^{01}(\check{\bf s}_R) \right],
\end{equation}
\begin{equation}
\frac{1}{2} \bigl[
\left(R_{TT}^{\prime t} {\cal S}_t + R_{TT}^{\prime l} {\cal S}_l \right)
\bigr]  =
i \left[\overline{W}^{12}(\check{\bf s}_R)-
\overline{W}^{21}(\check{\bf s}_R) \right].
\label{eq:rttp}
\end{equation}
\narrowtext
Since the response functions, as defined above, are independent of $\beta$,
certain
restrictions can be imposed using simple kinematical arguments.  In particular,
in
``parallel kinematics'' when $\vec p = \vec q$, all of the response functions
must
vanish except for $R_L$, $R_T$ and $R_{TT}^{\prime \, l}$, which are
unconstrained,
and the normal and transverse parts of $R_{LT}$ and $R'_{LT}$ which are
constrained
to be
opposites~\cite{PVO87,PVO89}.  We point out that the constraints of parallel
kinematics have been violated in the work of Allder~\cite{AA92}.
Refs.~\cite{PVO87} and~\cite{PVO89} show how these response functions can be
extracted
from the cross section.

Now, we leave the general theory of the
$({\vec e}, e' {\vec p})$ reaction and focus on the
upcoming CEBAF experiment~\cite{CEBAF}.
Saha and
collaborators plan
on measuring the unpolarized cross section and the normal component (to the
photonuclear scattering plane) of the
proton polarization.
In terms of the above notation, the unpolarized differential cross section
is~\cite{PVOW85}
\begin{eqnarray}
\sigma_0(0) &=& \frac{M_N |{\bf p}|}{(2\pi)^3}
\left( \frac{d\sigma}{d\Omega_{k'}} \right)_{Mott}
\bigl\{ V_L R_L + V_T R_T \nonumber\\
&& \qquad + V_{TT} R_{TT} \cos 2\beta + V_{LT} R_{LT}
\sin\beta \bigr\},
\label{eq:s0}
\end{eqnarray}
where the subscript $0$ on $\sigma$ signifies that electron
helicity is averaged and the argument of $0$ implies that the
direction of the ejected proton
spin is averaged over.  The normal component of the polarization vector is
then defined as
\begin{eqnarray}
P_n^0 &=& \frac{1}{\sigma_0(0)} \frac{M_N |{\bf p}|}{(2\pi)^3}
\left( \frac{d\sigma}{d\Omega_{k'}} \right)_{Mott}
\bigl\{ V_L R_L^n + V_T R_T^n \nonumber\\
&& \qquad + V_{TT} R_{TT}^n \cos 2\beta +
V_{LT} R_{LT}^n \sin \beta \bigr\}.
\label{eq:pn}
\end{eqnarray}
Since the normal component of the polarization, $P_n^0$, is dependent only
on normal response functions which vanish in the high energy of the CT limit,
$P_n^0$ also vanishes in
that limit.  Parity and time reversal invariance imply that non-zero values
of $P^0_n$ arise only from the final-state spin-orbit interaction.  If CT
is obtained {\it all} final-state interactions vanish, so that $P^0_n$
vanishes too.

It is also useful to consider the ``total" unpolarized cross section,
obtained by integrating over the solid angle of the outgoing proton.
If we integrate over $\beta$ in
Eq.~(\ref{eq:s0}), all of the terms which
have explicit $\beta$ dependence
vanish.  It then remains to integrate over $\sin \zeta \, d\zeta$.
However, at this stage, we introduce the convenient approximation that
$d\Omega_p \approx d^2 |{\bf p} - {\bf q}|/|{\bf q}|^2$.
Since
the response functions are functions
only of ${\bf q}_\bot \equiv {\bf p}-{\bf q}$, ($q_\bot =
|{\bf q}_\bot|$)
we define the total unpolarized cross
section as
\begin{eqnarray}
\sigma & \equiv & \frac{ d^2 \sigma}{d\epsilon_{k'} d\Omega_{k'}} \\*
&=& \frac{M_N |{\bf p}|}{\left(2\pi\right)^2  |{\bf q}|^2}
\left( \frac{d\sigma}{d\Omega_{k'}} \right)_{Mott}
\nonumber\\ && \qquad \times \int dq_\bot \, q_\bot
\left\{ V_L R_L \left( q_\bot \right) +
V_T R_T \left( q_\bot \right) \right\}.
\label{eq:stot}
\end{eqnarray}

The above equations completely specify the relationship between the
response functions of the $(e,e' \vec p)$ reaction and the experimental
observables which will be measured in the near future.
The crucial quantity is the electromagnetic nuclear current matrix element
(NCME) of Eq.~(\ref{eq:jmu}).

\section{DWBA Formalism}
\label{ch.glauber2}
\subsection{Bound State}
\label{sec:bound2}

We now
discuss the model bound state used here.
In Ref.~\cite{GM93}, the ejected proton
was taken to lie initially in a non-relativistic one-particle
harmonic oscillator shell-model state.  Since
the present formalism is spin dependent and relativistic, it is
necessary to look for
a relativistic single-particle shell-model
state which is a four-component Dirac spinor.
There are several such
models~\cite{CS86,SW86}.  Here we use
the finite nuclei model of
Horowitz and Serot~\cite{HS81}.  Our opinion is that
QHD is not
a fundamental field theory. However,
for our purposes it is not necessary to consider QHD to be anything but
a model, which accounts for nuclear phenomenology.  In particular, QHD
yields realistic nuclear densities and
four-component bound state wavefunctions.

It is useful to display the solution to the Dirac equation of QHD in terms of
radial and angular solutions.
The set of quantum numbers $\alpha = \left\{ n,j,l,m,t
\right\}$ where $n$ is the principal quantum number, $j$ is the total
angular momentum, $l$ is the orbital angular momentum, $m$ is the
$\check{\bf Z}$ component of the total angular momentum and $t$ is the third
component of the isospin.  Then, the solution is
\begin{eqnarray}
\langle {\bf R} | \alpha \rangle & \equiv &
\Phi_\alpha({\bf R})
\nonumber\\ & = & \frac{1}{R}
\left( \begin{array}{c} i G_{njl}(R) \\ + F_{njl}(R)
\mbox{\boldmath $\sigma$} \cdot
\check{\bf R}
\end{array} \right) {\cal Y}_{jlm}(\check{\bf R}) \, \eta_t,
\label{eq:bssol}
\end{eqnarray}
where $G$ and $F$ are functions of the radial variable $R= |\bf R|$ only
and
$\eta_t$ is a two-component isospinor, with
$\eta_t=\frac{1}{2}$
for protons and
$\eta_t=-\frac{1}{2}$
for neutrons.  The quantities ${\cal Y}_{jlm}$ are the standard
spin spherical harmonics.
We use a
computer program
by Horowitz~\cite{HO}
to generate these radial wavefunctions.
Note also that $\Phi_\alpha({\bf R})$ are normalized to unity.

\subsection{Current Operator}
\label{sec:current2}
The vector current operator represents the absorption of a virtual photon
by a nucleon bound in a nucleus.  This nucleon is somewhat off the energy
shell.
In general the current operator can be expressed in terms of scalar
functions of the four-momenta
multiplied by any of {\it twelve} four-vectors~\cite{PVO89}!
Since the bound
nucleon is only slightly off shell (the binding energy is small), and since
any other choice for this operator is only a guess, the current operator
is chosen
to be the free nucleon current operator.
Thus, we define the configuration-space
matrix elements, which are matrices in the Dirac space, as,
\begin{equation}
\langle {\bf R} |
{T}_H^\mu (q) | {\bf R'} \rangle
= e^{i{\bf q} \cdot {\bf R}} \delta^3({\bf R} - {\bf R'}) \, \gamma^0
\Gamma^{\mu}(q),
\label{eq:thmu}
\end{equation}
where
\begin{equation}
\Gamma^\mu(q)  =
F_1(q^2) \gamma^\mu + \frac{i}{2M_N}
F_2(q^2) \sigma^{\mu\nu}q_\nu,
\label{eq:gammamunu}
\end{equation}
and $F_1(q^2)$ and $F_2(q^2)$ are the usual Dirac and Pauli
form factors of the
nucleon.
We use a simple dipole parameterization such that
\begin{eqnarray}
G_E(q^2) &=& \left( 1 - \frac{q^2}{0.71 \, GeV^2} \right)^{-2},
\label{eq:formge}\\
G_M(q^2) &=& (1 + \kappa) G_E(q^2),\label{eq:formgm}
\end{eqnarray}
where $\kappa = 1.79$ is the anomalous magnetic moment of the proton.
The Dirac and Pauli form factors are given by
\begin{eqnarray}
F_1(q^2) &=&
\frac{q^2G_M(q^2) - 4M_N^2 G_E(q^2)}{q^2-4M_N^2}
, \\
F_2(q^2) &=& \frac{4M_N^2}{q^2 - 4M_N^2} \left[ G_E(q^2) - G_M(q^2)
\right]
{}.
\end{eqnarray}
This assumption
of the current operator neglects the possibility that the nucleon
properties are much modified in medium.

\subsection{Optical Potential and Distorted Wave}
\label{sec:opdw2}
We now describe the wavefunction of the scattered
proton.
Here the bound state proton wavefunction is a
relativistic four-component object, and the current operator is a
4 $\times$ 4 matrix.  Thus the proton wave must also be a Dirac
spinor, satisfying a one-particle Dirac equation.

Treating the Dirac equation as a one-particle equation has many serious
diseases.  Indeed, problems of interpretation with the
negative energy states led to the development of quantum field theory.
 However, the difficulties can be identified
(see e.g. Section 2 of Ref.~\cite{IZ}.)
Here we are interested in the
scattering of energy eigenstates of an interacting Dirac Hamiltonian.
Such eigenstates
propagate along with no transitions to the negative energy states.  In these
situations, there is no problem in the interpretation of the Dirac equation
as the governing equation of a single particle relativistic quantum
mechanics.

In the early
eighties, a phenomenology based on using the Dirac equation to describe the
scattering of medium energy
protons from nuclei was developed.  Both partial wave
analyses~\cite{ACMS81,RSM82} and eikonal
reductions of the Dirac equation~\cite{APSM83,PASM83,PI85,CDG87}
have been performed.
The successes of these
models, most notably the spin observables such as analyzing powers and
spin
rotation functions, as well as differential cross sections,
are still quite impressive.
{}From the standpoint of CT, where we are
interested in very high energy scattering, the eikonal form is appropriate.

Let us now discuss the potentials that enter into the Dirac equation.
In the impulse approximation these are complex
Lorentz scalars and time-components of 4-vectors.
We label these $V_s$ and $V_v$,
respectively.
In the relativistic impulse approximation,
the scalar and vector optical potentials, $V_s$
and $V_v$, are proportional to forward Dirac scalar and Dirac vector
scattering amplitudes ($F_s^0$, $F_v^0$)
and to the scalar and vector nuclear
densities~\cite{SMW83,MRW83,MSW83,WA81,WA84,WA86,WA87}
($\rho_s$, $\rho_v$), such
that
\begin{equation}
U_{opt}({\bf R}) = V_s({\bf R}) + \gamma^0 V_v({\bf R}),
\label{eq:uoptvsvv}
\end{equation}
where
\begin{equation}
V_s({\bf R}) = r \, F_s^0 \rho_s({\bf R})
\quad \hbox{\rm and} \quad V_v({\bf R})= r \, F_v^0
\rho_v({\bf R}),
\label{eq:vsvv}
\end{equation}
and $r=-4\pi i p_{lab} / M_N$ is a kinematical factor needed
to relate the invariant Feynman amplitude to the usual description of
the optical potential in the impulse approximation.
The
scalar and vector nuclear densities are taken from the QHD model in the
previous section~\cite{HS81},
\begin{eqnarray}
\rho_s(R)&=&\sum_{\alpha} \left( \frac{2j_\alpha+1}{4\pi R^2}\right)
\left( |G_\alpha(R)|^2 - |F_\alpha(R)|^2 \right),\\
\rho_v(R)&=&\sum_{\alpha} \left( \frac{2j_\alpha+1}{4\pi R^2}\right)
\left( |G_\alpha(R)|^2 + |F_\alpha(R)|^2 \right).
\end{eqnarray}
The potential strengths, $F_s^0$ and $F_v^0$ are taken directly from
$NN$ scattering
data~\cite{WA81,SJW,HK69,HK76,HO76,HO78,HH79,MSWY81,MSY86};
see Appendix A.  Since the strengths
are a sensitive function of energy and NN scattering data, we
consider our calculations reliable only at the energies where sufficient
data exists to determine the strengths $F_s^0$ and $F_v^0$.

The Dirac equation for the
distorted wave is given by
\begin{eqnarray}
H \Psi_{{\bf p},{\check{\bf s}_R}}^{(+)} ({\bf R})
&=& \Bigl[ -i \mbox{\boldmath $\alpha$} \cdot \mbox{\boldmath $\nabla$}
+ \beta \left(M_N+V_s(R)\right)
\nonumber\\* && \qquad \qquad \qquad + V_v(R) \Bigr]
\Psi_{{\bf p},{\check{\bf s}_R}}^{(+)} ({\bf R})\nonumber\\
&=& E \Psi_{{\bf p},{\check{\bf s}_R}}^{(+)} ({\bf R}),
\label{eq:diraceq}
\end{eqnarray}
where $E=\sqrt{p^2+M_N^2}$. This equation is solved by separation
into two coupled first-order differential equations.
The eikonal form of the solution to this equation
is well known~\cite{APSM83,PASM83,PI85,CDG87} and is given by
\begin{eqnarray}
\Psi_{{\bf p}, -{\check{\bf s}_R}}^{(+)} ({\bf R}) & = &{\cal N}
\left( \begin{array}{cc}
1\\ \frac{-i
\mbox{\boldmath $\sigma$} \cdot \mbox{\boldmath $\nabla$}}{E+M_N+V_s(R)-V_v(R)}
\end{array} \right)
e^{ipZ}
\nonumber\\ && \quad \times
\exp \left[ \int_{-\infty}^Z dZ' \Omega ({\bf B},Z')
\right] \chi_{-{\check{\bf s}_R}} ,
\label{eq:ufam}
\end{eqnarray}
where ${\bf R} \equiv {\bf B} + Z\check{\bf Z}$,
\begin{eqnarray}
\Omega ({\bf B},Z') &=& \frac{1}{2ip}
\Bigl\{ U_c({\bf B},Z')
\nonumber\\* && \qquad
+ U_{so}({\bf B},Z')\left[
\mbox{\boldmath $\sigma$} \cdot
{\bf B} \times {\bf p} - ipZ'
\right] \Bigr\} ,
\end{eqnarray}
${\bf R} \cdot {\bf p} = pZ$, and ${\bf B} \cdot {\bf p} = 0$.  Also note
that we have
used outgoing boundary conditions and
changed the sign of the rest-frame spin projection, as in
Eq.~(\ref{eq:jmuout}).
The normalization is ${\cal N}=\sqrt{(E+M_N)/(2M_N)}$.
The Dirac scalar and vector potentials have been eliminated in favor
of new central ($U_c$) and spin-orbit ($U_{so}$) potentials:
\begin{eqnarray}
U_c(R) &=& 2EV_v(R) + 2M_NV_s(R)
\nonumber\\* && \qquad \qquad + V_s^2(R) - V_v^2(R),
\label{eq:uc}\\
U_{so}(R) & =& \frac{1}{E+M_N+V_s(R)-V_v(R)}
\nonumber\\ && \qquad \qquad \times \frac{1}{R}
\frac{\partial}{\partial R} \left\{V_v(R)-V_s(R)\right\}.
\label{eq:uso}
\end{eqnarray}
Note that the potentials $U_c$ and $U_{so}$ have units of $(Energy)^2$;
we use the symbol $U$ (and not $V$) to indicate this.
The expression in Eq.~(\ref{eq:ufam}) can be simplified; the detailed form
of the wavefunctions is
relegated to Appendix B.

Finally, we note one other complication.  The eikonal wavefunctions
presented in Appendix~\ref{app:vectorctwf} have
been derived in the limit where the particle travels along the $\check{\bf Z}$
axis (direction of the virtual photon).   At very high energies,
the difference between $\check{\bf p}$ and $\check{\bf Z}$ is small.
N\"aively, this finite scattering angle should make little difference.
However, if one assumes that the scattering angle is exactly zero, then the
response function $R_{TT}$ is also exactly zero.  Since, in the plane-wave
limit, we know this is not true, account must be taken of this scattering
angle.  We have derived the cross section in Eq.~(\ref{eq:eepcs}) with the
coordinates that the final proton momentum makes an angle $\zeta$
with the $\check{\bf Z}$-axis (the direction of $\bf q$) and an angle
$\beta$ with normal to the electron scattering plane (the XZ plane).
Call this the
$q$-basis.  The derivation of the eikonal wavefunction described above has
the proton momentum $\bf p$ along $\check{\bf Z}$;  call this the $p$-basis.
To move our eikonal wavefunctions from the $p$-basis to the $q$-basis, we
perform a passive rotation of the coordinates  to the desired frame.

\section{Inclusion of CT}
\label{ch:incvct}
To include the effects of color transparency, we proceed as in
Ref.~\cite{GM93} and include the baryonic internal degrees of freedom.
Thus, the states in the full Hilbert space now carry two labels: one for
the ``external'' nuclear space and one for the internal quark space.
The quark space
operators are identified wearing ``hats'' and we write CT quantities using
script lettering,
where possible.  Then, the CT scattering matrix
element is given by
\begin{equation}
{\cal M}_\alpha = j_\mu {\cal J}^\mu_\alpha
\label{eq:mjjvct}
\end{equation}
where the electron current is $j_\mu$.
The triple differential cross section is given by
Eq.~(\ref{eq:crossv}), except for the replacement of $M_\alpha$
by ${\cal M}_\alpha$.  Thus, in order to completely specify
the method for the inclusion of CT effects, it remains to
construct the CT generalized NCME.  We
follow the method in Ref.~\cite{GM93} and write
\begin{equation}
{\cal J}^\mu_\alpha(q)
= \langle N, \alpha | {\widehat T}_H^{\mu \dagger}(q) | \Psi_{N,
{\bf p}, -\check{\bf s}_R}
\rangle^{(+)},
\label{eq:genjmuout}
\end{equation}
Here, $|\Psi_{N,{\bf p}, -{\check{\bf s}_R}} \rangle^{(+)}$
is a vector in both internal and external spaces.
The subscript on $\Psi$ is to indicate that ultimately a nucleon $N$
moving with momentum $\bf p$ and rest-frame spin projection
$-{\check{\bf s}_R}$ is detected.  The overlap of
this state with the nuclear position
state $\langle {\bf R} |$ gives a vector in
the internal space only, which we identify with bold face type:
$\langle {\bf R} | \Psi_{N, {\bf p},
-{\check{\bf s}_R}} \rangle^{(+)} = {\bf \Psi}_{N, {\bf p},
-{\check{\bf s}_R}}^{(+)} ({\bf R}) $.
The asymptotic boundary condition is implemented by imagining that
${\bf \Psi}_{N, {\bf p}, -{\check{\bf s}_R}}$
is some internal space vector, whose equation of motion we shall derive
below, projected onto the nucleon state $|N \rangle$.

In our approach
the relativistic bound state is not changed by including
CT effects.  See, however, Ref.~\cite{FSZ93}.  The rest of this Section
describes the other elements in
Eq.~(\ref{eq:genjmuout}).

\subsection{Current Operator}
\label{sec:currentct2}
Upon absorption of a high-energy photon, the proton is converted into a
coherent superposition of baryon states (a wavepacket),
with components labelled by
a discrete quantum number, $m$.  Then, we identify the relevant inelastic
transition Dirac and Pauli form factors
by taking the matrix elements of the current operator to
be
\begin{equation}
\langle N, {\bf R} |
{\widehat T}_H^\mu (q) | m, {\bf R'} \rangle
= e^{i{\bf q} \cdot {\bf R}} \delta^3({\bf R} - {\bf R'}) \, \gamma^0
\Gamma^{\mu}_{N,m}(q),
\label{eq:thmuct}
\end{equation}
where
\begin{equation}
\Gamma^{\mu}_{N,m}(q) =
F_{1_{N,m}}(q^2) \gamma^\mu + \frac{i}{2M_N}
F_{2_{N,m}}(q^2) \sigma^{\mu\nu}q_\nu.
\label{eq:gammamunumn}
\end{equation}

\subsection{Optical Potential}
\label{sec:optical2}
We now consider the
optical potentials $V_s(R)$ and $V_v(R)$
to be operators in the internal quark space.
These optical potentials are essentially products of the
baryon-nucleon scattering amplitude operator $\widehat f(\widehat b^2)$
with $\rho_s$ or $\rho_v$.
We choose the scalar and vector potentials to have the same
operator structure, to
be proportional to the same function $f({\widehat b}^2)$.
At present, there is no detailed knowledge about the precise form for
$f({\widehat b}^2)$.  However, some general properties are known.  For small
wavepackets
with $b^2 \ll b_H^2$ with $b_H^2 \equiv \langle N | {\widehat b}^2 |
N \rangle$, $f$ should vanish.  Interactions do occur for larger
wavepackets.  For non-zero but still small sized wavepackets, the
interaction goes like
\begin{equation}
\lim_{{\widehat b}^2 \rightarrow 0}
f({\widehat b}^2) \rightarrow \frac{{\widehat b}^2}{b_H^2}.
\label{eq:fsmall}
\end{equation}
The operator $f$ is normalized such that
\begin{eqnarray}
f({\widehat b}^2 = 0 ) &=& 0, \\
\langle N | f({\widehat b}^2) | N \rangle &=& 1.
\label{eq:fnorm}
\end{eqnarray}
We use a general function $f({\widehat b}^2)$, subject only to the
constraints of Eqs.~(\ref{eq:fsmall})-(\ref{eq:fnorm}).
The matrix elements of the scalar and
vector operators can then be written as
\begin{equation}
\langle m, {\bf R} | {\widehat V}_{s(v)} | m', {\bf R'} \rangle =
\delta^3 \! ({\bf R} - {\bf R'}) V_{s(v)}({\bf R})
\langle m | f({\widehat b}^2) |
m' \rangle.
\label{eq:vsvvct}
\end{equation}
To compactify the notation, we define the quantity
$V_{s(v)}({\bf R}) f({\widehat b}^2) \equiv {\widehat V}_{s(v)}({\bf R})$.

\subsection{Distorted Wave}
\label{sec:dw2}
In this section we obtain the wave equation for the
propagation of the wavepacket formed in the hard
collision through the nucleus.
Start by considering the time-independent Dirac equation.
The equivalent Dirac Hamiltonian should now be considered
an operator in the internal quark space.  As such, we award $H$ a
``hat''.  All of the tools have already been developed and we
immediately write
\begin{eqnarray}
{\widehat H }
{\bf \Psi}_{N,{\bf p},{\check{\bf s}_R}}^{(+)} ({\bf R})
&=& \Bigl[ -i \mbox{\boldmath $\alpha$} \cdot \mbox{\boldmath $\nabla$}
+ \beta \left({\widehat M}+{\widehat V}_s(R)\right)
\nonumber\\* && \qquad \qquad \qquad + {\widehat V}_v(R) \Bigr]
{\bf \Psi}_{N,{\bf p},{\check{\bf s}_R}}^{(+)} ({\bf R})\nonumber\\*
&=& E {\bf \Psi}_{N,{\bf p},{\check{\bf s}_R}}^{(+)} ({\bf R}),
\label{eq:diraceqct}
\end{eqnarray}
where $E$ is the energy and we have already taken the matrix element in the
external configuration space.
That is, ${\bf \Psi}_{N,{\bf p},{\check{\bf s}_R}}^{(+)}
({\bf R})
= \langle {\bf R} | \Psi_{N,{\bf p},{\check{\bf s}_R}}\rangle^{(+)}$.
In this notation, ${\widehat M}^2$ is the baryon mass operator squared.
That is,
${\widehat M}^2 | m \rangle = M_m^2 |m\rangle$; the
nucleon is the ground state with $m=N$  having eigenvalue $M_N^2$.
The combination
\begin{equation}
{\widehat{\bf p}}^2 \equiv E^2 - {\widehat M}^2
\end{equation}
is also a quark space operator which can be interpreted as the baryon
momentum operator squared.  This operator
 accounts for the
different kinematics with which the wavepacket components propagate through
the nucleus.  Here ${\bf {\widehat p}}$ is also a three-vector
in the
$\check{\bf Z}$ direction; ${\bf {\widehat p}} = {\widehat p}
\check{\bf Z}$.

The Dirac equation can again be solved by eliminating the lower components.
Then
we see that the CT scalar and vector Dirac potentials (which are operators
in the internal space) can be eliminated in favor of combinations
given by
\begin{eqnarray}
{\widehat U}_c(R) & = &
2E {\widehat V}_v(R) + \left\{{\widehat M},{\widehat V}_s(R)\right\}
+ \left[{\widehat M},{\widehat V}_v(R)\right]
\nonumber\\ && \qquad \qquad \qquad +
{\widehat V}_s^2(R) - {\widehat V}_v^2(R),
\label{eq:defuc}\\
{\widehat U}_{so}(R) &=& \frac{1}{R} \left[ \frac{\partial}{\partial R}
\left( {\widehat V}_v(R) - {\widehat V}_s(R) \right) \right]
\nonumber\\ && \qquad \times
\left( E+{\widehat M} + {\widehat V}_s(R)
- {\widehat V}_v(R) \right)^{-1},\label{eq:defuso}
\end{eqnarray}
where
the square (curly) brackets are the (anti-) commutator symbols.
We also define
``path-evolved'' versions of the operators
of Eqs.~(\ref{eq:defuc}) and~(\ref{eq:defuso}) as
\begin{equation}
{\widehat{\cal U}}_{c(so)}({\bf R})
=  e^{-i{\widehat p} Z} \frac{1}{{\widehat p}}
{\widehat U}_{c(so)}(R) e^{i {\widehat p} Z}.
\label{eq:uhatint}
\end{equation}

Eikonalizing in the usual way, we see that
the solution to the resulting eikonal first-order equation
is a path-ordered exponential.
The notation is simplified by defining an
operator ${\widehat \Omega}$ such that
\begin{eqnarray}
{{\widehat \Omega}}({\bf B},Z') &=& \frac{1}{2i} \biggl[
{\widehat{\cal U}}_c({\bf B},Z')
\nonumber\\* && \qquad
+ {\widehat{\cal U}}_{so}({\bf B},Z')
\left( \mbox{\boldmath $\sigma$} \cdot {\bf B} \times {\bf {\widehat p}} -
i {\widehat p} Z' \right)
\biggr].
\label{eq:defw}
\end{eqnarray}
Note that
$\left[ {{\widehat \Omega}}({\bf R}),{{\widehat \Omega}}({\bf R'}) \right]
\ne 0$ since ${\widehat p}$
and ${\widehat b}$ do not commute.
This is because ${\widehat p}$ contains
${\widehat M}^2$,
and $\left[ {\widehat M}^2, f({\widehat b}^2) \right] \ne 0$.

We want to identify the equivalent CT wavefunctions which are obtained
when the NCME is written as
\begin{equation}
{\cal J}_\alpha^\mu(q) = \int d^3 \! R \, \overline{\Phi}_\alpha({\bf R}) \,
e^{-i {\bf q} \cdot {\bf R}} \,
{\overline \Gamma}^\mu (q) \,
\Psi_{CT,{\bf p},-{\check{\bf s}_R}}^{(+)}({\bf R}).
\label{eq:xjmuct}
\end{equation}
where
$\overline{\Phi}_\alpha({\bf R}) = \Phi_\alpha^\dagger({\bf R}) \gamma^0$
and $\overline{\Gamma}^\mu = \gamma^0 \left[ \Gamma^\mu \right]^\dagger
\gamma^0$.  The
bound state, $\Phi_\alpha({\bf R})$, is given in Eq.~(\ref{eq:bssol}) and the
nucleon current operator is defined in Eq.~(\ref{eq:thmu}).
Quite generally, then, the CT wavefunction is given by
\widetext
\begin{eqnarray}
\Psi_{CT,{\bf p},-{\check{\bf s}_R}}^{(+)}({\bf R}) &=& {\cal N} \sum_{m}
\left[
\overline{\Gamma}^\mu (q) \right]^{-1}
\overline{\Gamma}^\mu_{N,m}(q) \qquad \times
\nonumber \\*
&& \! \! \! \! \! \! \! \! \! \! \! \! \! \!  \! \! \! \! \! \! \!
\! \! \! \! \! \! \! \! \!
\langle m | \left( \begin{array}{c}
1 \\ \frac{-i
\mbox{\boldmath $\sigma$} \cdot \mbox{\boldmath $\nabla$}}{E+{\widehat M}+
{\widehat V}_s({\bf R}) - {\widehat V}_v({\bf R})} \end{array} \right)
e^{i {\widehat p} Z} {\cal P}
\exp \left[ \int_{-\infty}^Z dZ' \, {\widehat \Omega}({\bf B},Z')
\right] |N\rangle \chi_{-{\check{\bf s}_R}},
\label{eq:ctwf}
\end{eqnarray}
\narrowtext
where we have taken the momentum of the outgoing wavepacket to lie along
the $\check{\bf Z}$ direction in Eq.~(\ref{eq:ctwf}) but not in the
calculations.  Again, ${\cal N} = \sqrt{(E+M_N)/(2M_N)}$.
No sum over the index $\mu$ is implied in the
above equation.
Above, $\cal P$ is the path-ordering symbol.  The effects of CT effects can
be seen by
comparing this expression with Eq.~(\ref{eq:ufam}).
We have projected the column vector $\bf \Psi$ on to the
nucleon state to be detected, and could also project
on to a nucleonic isobar to obtain the isobar production amplitude.
The inelastic nucleon current operator, $\Gamma^\mu_{N,m}$, is defined in
Eq.~(\ref{eq:thmuct}).

The expression for the CT wavefunction, shown in
Eq.~(\ref{eq:ctwf}), is the central formal result of this paper.
Note that we have changed the direction of the rest-frame spin projection,
in accordance with Eq.~(\ref{eq:genjmuout}).

\subsubsection{Approximations: Vector OBO, LEE and EA}
\label{sub:vapprox}

At this stage, we may proceed to make the
expansions and approximations, as in Ref.~\cite{GM93}, which
allow a numerical evaluation of the CT wavefunctions,
Eq.~(\ref{eq:ctwf}), the NCME, Eq.~(\ref{eq:xjmuct}) and
ultimately of the $(e,e' \vec p)$ cross section.  The definitions of these
approximations are only slight generalizations of those in Ref.~\cite{GM93}.
The ``Order by Order'' ($OBO$) scheme
is defined by simply expanding the path-ordered exponential
in Eq.~(\ref{eq:ctwf}) order
by order in its Taylor series and evaluating the resulting matrix elements
between the $PLC$ and the nucleon.  The Low-energy Expansion
($LEE$) scheme is defined by
factoring out the
expectation value of this operator ${\widehat \Omega}$ in the nucleon
by letting
${\widehat \Omega}({\bf R})
= \langle N | {\widehat \Omega}({\bf R}) |N\rangle
+ \Delta {\widehat \Omega}({\bf R})$
and treating $\Delta {\widehat \Omega}$ as a perturbation.
One must be a little
careful, because of the presence of the noncommuting Pauli
matrices.  We have also calculated the
``Exponential Approximation'' ($EA$) defined by
exponentiating the first-order result of the OBO (this is equivalent to
neglecting the path ordering).
This is similar to letting
$1-x \approx e^{-x}$.  Our previous experience leads us to believe that the
$EA$ and the $LEE$ will be accurate at the low and medium energies while
the $EA$ and the $OBO$ should be reliable at the highest energies.
This is verified below.

\section{Model Evaluations and Applications}
\label{ch:eval2}

\subsection{Wavepacket-Nucleon Interaction and Quark Space}
\label{sec:wpniqs2}

To proceed further we need specific forms for $\widehat f$ and the states $m$.
We choose the
interaction to be
\begin{equation}
f({\widehat b}^2) = {\widehat b}^2/b_H^2,
\end{equation}
as in Eq.~(\ref{eq:fsmall}).
Further,
we choose the internal baryonic states to
be described by a two-dimensional transverse harmonic oscillator.
For interactions of the form $f(\widehat b^2)$, the use of three
dimensional oscillators leads to the same results as those of the
two-dimensional case.
We calculate transparencies for  $b_H = 1 \, fm$ and for two different
values of the oscillator spacing:
$M_2^2-M_N^2 = 1.19 \, GeV^2$ and $M_2^2-M_N^2 = 2.36 \, GeV^2$.
These simple choices reasonably represent the present (lack of detailed)
knowledge of $f(\hat b^2)$ and the baryon wavefunctions.
With the internal model space and interaction operator specified, we
note the useful formula:
\begin{eqnarray}
\langle 2m | \frac{{\widehat b}^2}{b_H^2} | 2n \rangle
&=& (2m+1)\, \delta_{m,n} - m \,
\delta_{m,n+1} \nonumber\\ && \qquad \qquad - (m+1) \, \delta_{m+1,n}.
\label{eq:matel}
\end{eqnarray}

Before we describe our evaluations, it is necessary to
show the parameters we use for the strengths of the optical potentials.
These are determined in
Appendix~\ref{app:nnamp}  and displayed
in Table~\ref{tab:strengths}.  It is useful, when examining the
table, to recall Eqs.~(\ref{eq:uoptvsvv}) and~(\ref{eq:vsvv}).  Also,
recall Eqs.~(\ref{eq:uc}), (\ref{eq:uso}), (\ref{eq:defuc}) and
(\ref{eq:defuso}) to see that at high energies the central potential
dominates and that, within the central potential, the vector potential is
the important quantity.  Thus, we can immediately see that at high energies
 the optical potentials we display in Table~\ref{tab:strengths} are
absorptive as the usual optical potentials.

The last column of Table~\ref{tab:strengths} shows the potential strengths
in nuclear matter in more conventional units, where we have taken
$\rho_0= 0.166 \, fm^{-3}$.  The optical potential
strengths are energy dependent, but this energy dependence is unknown
for higher energies where there is no data.
Therefore,
we have no way to really know what the strengths should be at higher
energies.  The strengths are totally determined by NN
scattering data, but they are functions of more than just the
invariant cross sections, see Appendix~\ref{app:nnamp}.  Thus,
{\em we only
consider our calculation to be reliable at energies where enough data
exists to at least allow an approximate determination of the strengths}.

As described in Appendix~\ref{app:nnamp}, the strengths displayed in
Table~\ref{tab:strengths} can be arranged to give the proton-nucleon
total cross section
and the ratio of the real to imaginary parts of the forward scattering
amplitude, $\alpha_f$.  These numbers are summarized in Table~\ref{tab:csr}.

Defining some more notation helps us proceed.  First, we
introduce the shorthand notation $U^{(i,j)}$ such that
\begin{eqnarray}
\langle 2m |
{\widehat U}_c({R}) | 2j \rangle & \equiv &
U_c^{(2m,2j)}({R})
\\
\langle 2m | {\widehat U}_{so}({R}) | 2j \rangle & \equiv &
U_{so}^{(2m,2j)}({R}).
\end{eqnarray}
In the above equation,
it is understood that $U_c^{(2j \pm 2n, 2j)}$ vanishes for
$n \ge 3$ while
$U_{so}^{(2j \pm 2n , 2j)} \equiv 0$ for $n \ge 2$.

More explicitly, we can evaluate the matrix elements just defined in terms
of the scalar and vector optical potentials, using Eq.~(\ref{eq:matel}).
Evaluation of the spin-orbit potentials require a little more discussion.
In particular, the inverse operator appearing in Eq.~(\ref{eq:defuso})
is problematical because
$\left[ {\widehat M}, {\widehat V}_{s(v)} \right] \ne 0$.  However,
neglecting the operators
${\widehat V}_s({\bf R})$ and ${\widehat V}_v({\bf R})$
in this inverse operator
is of the same order of approximation as those already made in deriving the
eikonal approximation so that it is safe to replace
\begin{equation}
\left[E+{\widehat M}+{\widehat V}_s({\bf R})-
{\widehat V}_v({\bf R})\right]^{-1} \approx
\left[E+{\widehat M}\right]^{-1},
\end{equation}
also in Eq.~(\ref{eq:ctwf}) to be consistent.
In that case, we can obtain explicit representations for the spin-orbit
CT potentials.  As an example,
\begin{eqnarray}
U_{so}^{(0,0)}({R}) &=& \frac{1}{R} \frac{\partial}{\partial R}
\frac{V_v({R})-V_s({R})}{E+M_N},
\label{eq:uso00}\\
U_{so}^{(2,0)}({R}) &=& -\frac{1}{R} \frac{\partial}{\partial R}
\frac{V_v(R)-V_s(R)}{E+M_2}.
\label{eq:uso20}
\end{eqnarray}
The first few central CT potentials have the
explicit representations
\begin{eqnarray}
U_c^{(0,0)}({R}) & = &
2EV_v({R}) + 2 M_NV_s({R})
\nonumber\\* && \qquad \qquad \qquad + 2 V_s^2({R}) - 2 V_v^2({R}),
\label{eq:uc0}\\
U_c^{(2,0)}({R}) & = &
\left( M_N-M_2-2E \right) V_v({R})
\nonumber\\* &&
\! \! \! \! \! \! \! \! \! \! \! \! \! \! \! \! \! \! \! \! \!
- \left( M_N + M_2
\right) V_s({R})
- 4 V_s^2({R}) + 4 V_v^2({R}), \label{eq:uc2}\\
U_c^{(4,0)}({R}) & = & 2 V_s^2({R}) - 2V_v^2({R}).\label{eq:uc4}
\end{eqnarray}
where we have only displayed the nonzero CT potentials which are connected
to the ground state (nucleon).  Since only the spin-orbit terms enter with
the spin operator, it is convenient to define
\begin{equation}
U_{cso}^{(2m,2j)}({\bf R}) = U_c^{(2m,2j)}({R}) - i p Z \,
U_{so}^{(2m,2j)}({R}).
\end{equation}

It is also
useful to define, here, a commonly appearing combination of the
above CT potentials:
\begin{equation}
{\cal U}_{c(so)}^{(2j)}({\bf B},Z,Z')  =  \sum_{m=0}^\infty
U_{c(so)}^{(2m,2j)}({\bf B},Z) e^{i(p_{2m}-p)(Z-Z')},
\end{equation}
In per\-form\-ing the cal\-cu\-la\-tions, one also encounters terms like
$\frac{p}{p_{2m}}$
and $\frac{E+M_N}{E+M_{2m}}$ which appear inside the summation sign
above.
In the spirit of the
high-energy nature and approximations of this paper, in the calculations
we assume all of these quantities to be equal.  Thus, the expressions
appearing in Eq.~(\ref{eq:uso00}) and Eq.~(\ref{eq:uso20}) are opposites
in this approximation.
The differences introduced by these approximations are
smaller than the other uncertainties in the calculation.

\subsection{Dirac Effects}
\label{sec:de}

We point out some interesting results obtained with this
Dirac formalism.  This is included in anticipation of the numerical
results discussed below.
The following effects are
general features of a relativistic description
of the wavepacket.

1) The presence of the
commutator term in
Eq.~(\ref{eq:defuc}) causes a new non-Hermiticity in ${\widehat U}_c({R})$.
Thus, care must be taken in order to obtain the correct sign of
the commutator term.  Because of the quark operator
structure present in
${\widehat U}_c$, and in ${\widehat U}_{so}$, it is a little bit tricky to
switch between outgoing wave and incoming wave boundary conditions.  In the
DWBA, all we do is take the complex conjugate of the potentials.
 Here, that will not suffice.  One must take the complex conjugate of the
strengths, while leaving the CT part alone, except for making the change in
this commutator term.  Because this switch is complicated and relatively
subtle, we only consider outgoing wave boundary conditions.
We have checked, however, that both descriptions give the same result.
Hereafter, this equality is to be understood and we drop the superscripts
indicating the boundary condition.

2) The presence of the quadratic terms in the central potential causes
an inequality between quantities
$U_c^{(0)}({R})$, as defined in Eq.~(\ref{eq:defuc}),
and $U_c({R})$ as defined in Eq.~(\ref{eq:uc}).  This effect
is an artifact of the way we solve the Dirac eikonal
formalism.

3) The commutator term in
${\widehat U}_c$ alters the approach to full transparency.
This is seen in the following way.  In the limit of zero size wavepacket,
the inelastic and elastic form factors are equal.  Then, we can neglect the
subscripts on $F_1$, $F_2$ and $\Gamma^{\mu}$.  In that case,
the initial $PLC$ is described simply
by $|PLC \rangle = {\widehat T}_H^\mu(q) |N
\rangle = \Gamma^{\mu}(q) \sum_m  | m \rangle$.  Then,
\begin{eqnarray}
\langle PLC | {\widehat U}_c({R}) | N \rangle &=& \Gamma^{\mu}(q)
 \sum_m \langle m |{\widehat U}_c({R}) | N \rangle, \\
&=& \Gamma^{\mu}(q)
\Delta M
\left(V_s({R}) + V_v({R}) \right),\label{eq:ctlim}
\end{eqnarray}
where $\Delta M = M_N-M_2$.
The reader is urged not to panic, however, despite the appearance that in
the closure limit CT is not obtained.  Upon looking at the definition of
the CT potentials which actually enter, Eq.~(\ref{eq:uhatint}), we see that
${\widehat U}_c$
always enters with a factor of ${\widehat p}$ in the denominator.  Thus, in
the high-energy limit
(closure limit) ${\widehat p} \approx p \rightarrow \infty$,
the exponential factors do not matter, and the $\frac{1}{p}$ factor will
suppress Eq.~(\ref{eq:ctlim}).  This is not unlike the situation in
Ref.~\cite{GM93}.  In that case, the approach to transparency was
described as $1-e^{i(p_2-p)\Delta} \sim (p_2-p) \Delta$ for large energies.
Since ${\widehat p} \approx p+ ( M_N^2 - {\widehat M}^2 )/(2p)$,
we see that $p_2-p \sim
(M_N+M_2)(M_N-M_2)/p$.  So, we see that CT is still obtained at very high
energies, only its approach is affected.  Furthermore, the terms which are
left over, in the CT limit, are of the same order as terms like
$p / p_{2j}$ which we have neglected already.

4) New color transparency derivative (CTD) effects enter.
To illustrate this effect,
consider the one-dimensional functions
\begin{eqnarray}
F(Z) &=& \int_{-\infty}^Z dZ' \, f(Z'),\\
F_{CT} (Z) &=& \int_{-\infty}^Z dZ' \, f(Z')
\left[ 1 - e^{i(p_2-p)(Z-Z')} \right].
\end{eqnarray}
If the energy in this fictitious problem is low enough so that the momentum
$p_2$ is below threshhold and thus purely imaginary, then it is
true that
\begin{equation}
F_{CT}(Z) \stackrel{E\rightarrow M^+}{\rightarrow} F(Z),
\end{equation}
assuming that $Z-Z'$ is ``large enough'' to kill the exponential.
Now, if we take the derivative with respect to $Z$ we see that
\begin{eqnarray}
F'(Z) &=& f(Z),\\
F'_{CT} (Z) &=& 0,
\end{eqnarray}
in the low-energy regime where the exponential will be damped because we
are below threshhold.  So, we are faced with the very interesting example
of two functions which approach each other as $E\rightarrow M^+$ but whose
derivatives do not.  These CTD effects are relevant for the Dirac formalism
used here because of the derivative
($ \mbox{\boldmath $\sigma$} \cdot \mbox{\boldmath $\nabla$}$)
which appears in the
lower component of the Dirac wavefunctions.  This effect is small

5) The Dirac strengths
have a significant real part.  The reader may be confused by the difference
between our approach and the usual procedure of using
the optical
theorem to relate the strength of the optical potential to the cross
section.  (There the real part of the scattering amplitude causes negligible
effects.)  Here
we convert the usual scattering amplitudes
into their invariant Dirac form to extract the strengths of the scalar and
vector potentials.  Even when the real part of the
forward scattering amplitude is neglected, the real parts of the Dirac
strengths do not vanish.  To illustrate this effect, we define a
quantity
\begin{eqnarray}
\Omega_v({\bf R}) &=& -i\int_{-\infty}^Z
dZ'\left(F_{v,r}^0 - i F_{v,i}^0
\right) \rho_v({\bf B},Z')
\nonumber\\* && \qquad \qquad \times \left[ 1 - e^{i(p_2-p)(Z-Z')} \right]
\end{eqnarray}
as a typical term in the CT equations which will dominate at high
energies.
Let us further define the integrals
above to be
\begin{eqnarray}
C_R({\bf B},Z) &=& \int_{-\infty}^Z dZ'
\rho_v({\bf B},Z')
\nonumber\\* && \qquad \qquad \times
\left[ 1 - \cos (p_2-p)(Z-Z')\right],\\
S_R({\bf B},Z) &=& \int_{-\infty}^Z dZ' \rho_v({\bf B},Z') \sin (p_2-p)(Z-Z'),
\end{eqnarray}
so that
\begin{eqnarray}
\Omega_v({\bf R}) &=& -F_{v,i}^0 C_R({\bf R}) - F_{v,r}^0 S_R({\bf R})
\nonumber\\* && \qquad +
i F_{v,i}^0 S_R({\bf R}) - i F_{v,r}^0 C_R({\bf R}).
\end{eqnarray}
Now, in the previous approach, it was as if $F_{v,r}^0$ were zero.  In that
case, since $C_R$ is positive definite, $e^{|\Omega_v|^2}$
was less than unity.
Because our strengths have a significant real part and because
$S_R$ is not positive definite, it is possible to obtain
$e^{|\Omega_v|^2} > 1$.  Therefore, transparencies greater than unity are not
disallowed.

The precise size of this nuclear enhancement effect
depends in detail upon the relative sizes and phases of
$C_R$ and $S_R$ as well as the strengths of the optical potentials.
Thus,  we cannot predict for certain that the
cross section ratios will exceed unity at presently feasible energies,
although we believe, on the grounds of the above argument, that this must
eventually happen.

\subsection{Zero Size}

We also assume that the hard
interaction forms a PLC. Then,
\begin{eqnarray}
\langle N | {\widehat T}_H^\mu(q) &=& \sum_m \Gamma^{\mu}_{N,m}(q) \langle m |,
\\
&=& C^\mu (q) \langle {\bf b} = 0 |
\end{eqnarray}
where
$\Gamma^{\mu}_{N,m}(q)$ is defined in Eq.~(\ref{eq:gammamunumn}).
Using the specified model, all of the form factors are equal and we can
neglect the subscripts on $F_1$, $F_2$ and on $\Gamma^{\mu}$
($\Gamma^\mu \equiv \Gamma^\mu_{N,N}$).  Then,
$\left( \Gamma^\mu \right)^{-1} \Gamma^\mu_{N,m}  = 1$.
Later on we shall return to the case where the inelastic form
factors are not equal to the elastic ones.  In particular, in this Dirac
case, interesting effects arising from a possible different
$F_1$ and $F_2$ dependence on the
size of the wavepacket are possible.

It is  simple to obtain explicit
evaluations of Eq.~(\ref{eq:ctwf})
within the various approximation schemes we have described.  The results
are lengthy; the interested reader can find them in
Appendix B.  With explicit expressions for the
wavefunctions in hand, we use Eq.~(\ref{eq:xjmuct}) to calculate the
nuclear current matrix element, from which we can construct the nuclear
tensor and all of the observables.

\section{DISCUSSION OF NUMERICAL RESULTS}
\label{sec:dnr}
We have displayed the nuclear current matrix element (NCME),
in Eq.~(\ref{eq:xjmuct}), as a three-dimensional integral,
with another  integral to be performed in the
wavefunction itself, for each order in perturbation theory.
In the basis where
$d^3R = BdB \, dZ\, d\phi$, the $\phi$ integral can actually be done exactly
and analytically~\cite{AA92},  but the result is not useful.
Therefore, we simply perform all of the
necessary integrals numerically.  We must include 50 integration points
per integral,
using Gaussian quadrature to get stable numerical results.

The displayed results use kinematics closely related to that of the
proposed experiments~\cite{CEBAF,EEF}.  We do not use precisely those
kinematics because our optical potential strengths are only known at
energies where previous free
NN data has been taken.
Predictions for the upcoming experiments can be
readily obtained using simple interpolations.  These kinematics are
summarized in Table~\ref{tab:kin}.
For all of our calculations, we take the angle $\beta = \pi/2$ and consider
only in-plane scattering.  The quantities $E_i$ and $\theta_e$ denote the
initial electron energy and the electron scattering angle, respectively.

\subsubsection{Convergence of Exponential Approximation}
 From our previous experience in Ref.~\cite{GM93}, we learned that the
exponential approximation (EA) was a good approximation to the full
multiple-scattering series.  This was seen by examining the
higher order terms in the various approximation schemes and showing that
all of the schemes converged to the
EA in the energy regime where
they were expected to be valid.  The interpretation of the success of the
EA is that
the neglect of the path ordering includes many more higher-order
terms
than one would naively expect.  We do not expect this conclusion to
change simply by changing over to a Dirac description.  To be careful
we check to see if this remains true.

We calculate the total cross section ratios for
the various approximation schemes described above, {\em in the forward
direction only} to cut down on the computational time required.
This should be a sufficient test.
In Figures~\ref{fig:o2l} and~\ref{fig:o2h} we display
the second-order cross section calculations in the forward direction, for
two values of the internal (quark) oscillator spacing.  To avoid
distraction by the energy dependence of the elementary proton-proton
scattering amplitude, we fix the optical potential strengths at their
$Q^2=5.96 \, GeV^2$ values.  Further, we have calculated these results at
the values $Q^2 =$ 1, 2, 3, 4, 5, 6, 8, 10, 12, 14, 16, 20, 25 and
30 $GeV^2$.

We see that the exponential approximation (EA) is a good
approximation to the true answer.  At low energies, the $LEE_1$
follows the EA
from the lowest energies to $Q^2 \approx 7 \, GeV^2$ for a light excited
state mass and up to $Q^2 \approx 12 \, GeV^2$ for a heavy mass.
The second-order $LEE_2$ does even better than the $LEE_1$ at higher
energies and approaches the $EA$.  It is clear that if one were to keep
including higher orders in the $LEE$ expansion, we would improve the higher
energy results.  At lower energies, the $LEE_2$ does not really converge to
the $EA$ as well as $LEE_1$.  While one could argue that the agreement
between the $EA$ and the $LEE_1$ is accidental, we do not take this view.  We
believe that at the lower energy end, the $LEE_2$ is actually less accurate
than the $LEE_1$.  This is because
we set $p/p_{2m}$ equal to unity in our CT potentials, which is accurate at
high energies. In a
first-order calculation, because of the presence of the quadratic terms, we
have neglected terms like $(p/p_4-1)$.  In the second-order calculation, we
have ignored terms like $(p/p_8-1)$.  This kind of effect should be especially
sizable a low energies.  In order to do this calculation more carefully,
one should keep these ``momentum fraction'' terms.  However, these terms
are of the same order as terms which have been neglected in the eikonal
approximation itself.  Thus, to do a really careful job, one should also
include the first non-eikonal correction to the scattering equations.  We
do not do this here.  The only point about the $LEE_2$ we want to make is
that at low energies it is approximately equal to the $LEE_1$ and the $EA$
and at higher energies, it more closely approximates the $EA$ than does the
first-order calculation.

At the high-energy end, we see that the $OBO_1$ fails terribly, even
for $^{12}C$.  Of course, we expect the $OBO$ to be increasingly worse for
heavier nuclei where the propagation lengths are longer.  However, adding
in the second-order contribution to the $OBO$ we find very good agreement
for $Q^2 \ge 16 \, (21) \, GeV^2$ between the $OBO_2$ and the $EA$, for light
(heavy) excited state masses.

At low energies, we really believe the $LEE$ ( the $LEE_1$ in particular)
and at high energies the $OBO$ is defined to describe the correct physics.
What the results in Figures 3 and 4 tell us is
that the approximation schemes overlap so that we can have confidence that
the $EA$ is the correct answer from $Q^2 \approx 1 $ to $ 7 \, (12)\,
GeV^2$ and from $Q^2 \ge 16 \, (21)\, GeV^2$ for light (heavy) excited state
masses.  Although we have not calculated higher order terms
we take the position, that with our previous experience in
Ref.~\cite{GM93}, we can believe that the $EA$ is a good
approximation to the full multiple-scattering series over the whole energy
range.

\subsubsection{Integrated Unpolarized Cross Sections}
We next display results for the ratios of integrated unpolarized
cross sections.  That is, we calculate the total cross section, as in
Eq.~(\ref{eq:stot}), and divide by the Born cross section,
defined by using a Dirac plane wave for the outgoing proton
wavefunction in Eq.~(\ref{eq:genjmuout}).  See Figures 5, 6 and 7.

The solid
curves (circles) are the DWBA cross sections, normalized to the
respective Born cross
sections.  The dotdashed curves (diamonds) denote the cross sections,
including CT effects via the exponential approximation, for oscillator
spacings of $\Delta M^2 = 1.19 \, GeV^2$ ($M_2 = 1.44 \, GeV$).  The dashed
curves show the cross section ratios, including CT effects via the
exponential approximation, for an excited state mass of $M_2=1.80 \, GeV$.
See Table~\ref{tab:kin} for
the electron kinematics.

There are several noteworthy features in the figures.  The first is that the
optical potential strengths, which are calculated from free nucleon-nucleon
scattering data, naturally give a DWBA
cross section ratio which is large at
$Q^2 \approx 1 \, GeV^2$ and decreases sharply with energy before reaching
an asymptotic value.  This can be understood by noting that the $pp$ cross
section is only $30 \, mb$ at this energy.  Thus, we have here an
effect similar to the one of Frankfurt,
Strikman, and Zhalov~\cite{FSZ93}.
Recall, the NE-18 people took
measurements at $Q^2=$1,3,5 and 6.8 $GeV^2$.  With an excited state mass
parameter of $1.44 \, GeV$ the cross section at $Q^2 \approx 3 \, GeV^2$ is
the same as at $Q^2 \approx 1 GeV^2$, although at $Q^2\approx 5$ and $ 7 \,
GeV^2$ the cross section rises rapidly with energy.

We want to stress, at this point, that the precise value
of the oscillator parameter, which
controls the value of the excited state masses
in the CT models, is unknown except for the fact that is on the order of
hadronic mass differences.

Therefore, by simply increasing the value of the first
excited state mass, we can postpone the onset of
transparency to higher energies, see the
dashed curves.  The suppression of the transparency for an excited state
mass of $1.80 \, GeV^2$ may be too great to agree with the SLAC data at
$Q^2 \approx 3 \, GeV^2$.
Thus, it seems that an oscillator
spacing such that $M_2 \approx 1.6 \, GeV$ would
give a cross section ratio which would appear to be $Q^2$-independent if
one only looked at $Q^2 = 1, 3, 5, 6.8 \, GeV^2$.

The SLAC data are taken at kinematics slightly different
than what we use.
However, for cross section ratios, the
differences in the electron kinematics lead to essentially the same results.

That the SLAC experiment has seen CT is certainly possible, given the
energy dependence of the elementary $pp$ cross section and the optical
potential strengths.  However, before one can be absolutely sure, it is
desirable to take more data points and see what happens at, say,
$Q^2 \approx 2 \, GeV^2$.  It would also be nice to increase
$Q^2$ and really see a dramatic rise in the ratio $\sigma/\sigma^B$.

\subsubsection{Integrated Longitudinal and Transverse Responses}
\label{sub:rlrt}
It has long been known that the longitudinal response function is
suppressed relative to the transverse response in the quasi-elastic
region~\cite{DG84,ME84,ME85,KS85,FP87,HO88}.
The theoretical interpretation of
this quenching has been difficult and no consensus,
beyond the failure of naive applications of
traditional nuclear theory,
has yet been achieved.  In particular, it has been suggested that a
modification of the properties of the nucleon in medium or the presence of
non-nucleon degrees of freedom is responsible for
this suppression.  It has also been
suggested that relativistic effects may play a role.  Indeed,
Do Dang and Van
Gai~\cite{DG84} have shown that the presence of large scalar and vector
potentials in the nucleus may be responsible for this effect.  Since our
calculation is based on such a scalar-vector phenomenology, it is
legitimate to ask whether or not our calculation displays this quenching.

As the integrated cross sections only involve the integrated longitudinal
and transverse response functions, Eq.~(\ref{eq:stot}), it is useful to
examine the ratio of
\begin{equation}
\frac{ {\cal R}_L}{{\cal R}_T} =
\frac{ \int d\zeta \, \sin \zeta \, R_L (\zeta) }
     { \int d\zeta \, \sin \zeta \, R_T (\zeta) } \approx
\frac{ \int dq_\bot \, q_\bot \, R_L (q_\bot) }
     { \int dq_\bot \, q_\bot \, R_T (q_\bot) },
\label{eq:rlrt}
\end{equation}
where $\zeta$ is the angle between the virtual photon and the ejected
proton and $q_\bot=|{\bf p}-{\bf q}| \approx |{\bf q}| \sin \zeta$.
Since the response functions $R_L$ and $R_T$ go to zero rapidly for
$q_\bot \ge 1.5 \, fm^{-1}$, we take the limits of integration in
$q_\bot$ to run from zero to some number $q_\bot^{max} \approx 2 \,
fm^{-1}$.  The contribution to the angle integrated responses
${\cal R}_{L(T)}$ from $q_\bot$ greater than 2 $fm^{-1}$ is negligible.
Note that this integration is over angle and
{\em not} over the energy transfer.  Thus, our ratio ${\cal R}_L /
{\cal R}_T$ has very little to do with the integral involved in
the Coulomb sum rule.

If longitudinal quenching occurs, ${\cal R}_L / {\cal R}_T < 1$.
The dependence of this ratio as a
function of $Q^2$, for quasi-elastic kinematics, can be inferred from the
concept of $y-$scaling~\cite{FS88,WE75,WE80,CDA83,LO86,FLC84,DMDS90,SI88}.
In order to determine the $Q^2$ dependence, we
define the new reduced response functions $r_L$ and $r_T$ as
\begin{eqnarray}
{\cal R}_L({\bf q}, q_0) &=& \frac{A}{|{\bf q}|} {\overline G}_E^2(Q^2)
\left( \frac{|{\bf q}|^2}{Q^2} \right) r_L({\bf q}, q_0) ,
\label{eq:irl}\\
{\cal R}_T({\bf q}, q_0) &=& \frac{A}{|{\bf q}|} {\overline G}_M^2(Q^2)
\left( \frac{Q^2}{2M_N^2} \right) r_T({\bf q}, q_0),\label{eq:irt}
\end{eqnarray}
where ${\overline G}_E$ and ${\overline G}_M$
are the electric and magnetic form factors of the
nucleus, which are weighted sums of the electric and magnetic form factors
of the constituent protons and neutrons.  The statement of $y$-scaling
is that at large enough momentum transfers, the reduced response functions
$r_L$ and $r_T$ are functions only of the variable
\begin{equation}
y= -\frac{|{\bf q}|}{2} + \frac{M_N \, q_0}{|{\bf q}|}
\sqrt{ \left( \frac{|{\bf q}|^2}{Q^2} \right) \left( 1 + \frac{Q^2}{4M_N^2}
\right) }.
\label{eq:scalingy}
\end{equation}
The quantity $y$ is to be interpreted as the longitudinal
(parallel to $\bf q$) component of the
initial nucleon momentum (in Born approximation) when the transverse
(to $\bf q$) component is zero.  We note that many choices of the
scaling variable exist.
At the top of the quasi-elastic peak ($x = Q^2/(2M_Nq^0)=1$),
$y=0$ for all values of $Q^2$.  Thus, in this kinematical limit,
which is precisely where we have calculated, the functions
$r_L$ and $r_T$ are independent of $Q^2$ (when y-scaling is valid).
In principle, $r_L(y)$ and $r_T(y)$ should be equal.  Experiments have
found that although $r_L$ and $r_T$ seem to functions of $y$ only, they are
not the same function~\cite{FLC84}.
Thus, simply from looking at the remaining $Q^2$ dependence in
Eqs.~(\ref{eq:irl}) and~(\ref{eq:irt}), we see that
\begin{equation}
\frac{ {\cal R}_L }{ {\cal R}_T} \stackrel{Q^2 \rightarrow
\infty}{\rightarrow} {\rm constant},
\end{equation}
from above.  This is indeed what we see in our calculation.

In Figures 8, 9 and 10
we display the ratio ${\cal R}_L/{\cal R}_T$ as a function of $Q^2$
for $^{12}C$, $^{40}Ca$ and $^{208}Pb$.
As before the circles, joined by
solid lines, show this ratio for the DWBA calculation.  The dotdashed
curves (diamonds) and dashed (boxes) curves show this ratio for the
CT-included cases (via the EA) for the light mass ($M_2=1.44 \, GeV$) and
the heavy mass ($M_2=1.80\, GeV$) cases.  We also display the result for
the plane wave or Born case, which is shown in the figures as the fancy
boxes joined by the dotted curves.

At higher energies the CT-included ratios are slightly more
suppressed than the DWBA.  But only slightly.  Thus, even if a careful
Rosenbluth-type separation could be performed, the differences in
${\cal R}_L / {\cal R}_T$ predicted by the
Glauber and CT calculations are too small to experimentally
differentiate between the two.

\subsubsection{Gauge Invariance and Current Conservation}

We have mentioned earlier that our calculation violates gauge invariance
and current conservation (CC).  One of the main causes is that,
in our NCME, we treat the bound
state and scattered state as derived from different Hamiltonia and thus are
not members of the same complete set.  This is a typical problem for
distorted wave calculations of this type.
(For the initial- and final-state wavefunctions to be derived from the same
Hamiltonian is necessary but not sufficient to obtain gauge invariance.)
Gauge invariance is often
imposed artificially~\cite{PVO89,DEF82}.  In our view this is not the best
solution, since one must be very careful to not introduce unphysical
terms into the calculation.  It is true that for our
calculation to make any sense at all, the current must be at least
approximately conserved.  This subsection examines the question of CC.

There are many ways that one can consider quantifying the amount of current
conservation violation (CCV).  The nuclear current has four complex
components,
so directly looking at $q_\mu {\cal J}^\mu$ seems tedious.  Instead, we
consider the relevant elements of the nuclear tensor, $\overline{W}^{\mu\nu}$.
We would like to investigate how the CCV's change as a function of
$Q^2$.  Since $\overline{W}^{\mu \nu}$ is a function of angle ($q_\bot$), we
define
\begin{equation}
{\overline{\cal W}}^{\mu \nu} \equiv \frac{2\pi}{|{\bf q}|^2}
\int dq_\bot \, q_\bot \, \overline{W}^{\mu \nu}.
\end{equation}
where
the energy-integrated
nuclear tensor ${\overline W}^{\mu \nu}$ is defined by
Eq.~(\ref{eq:wbar}).  The quantity $q_\bot = |{\bf p} - {\bf q}| \approx
|{\bf q}| \sin \zeta$, where $\zeta$ is the angle between the
the virtual photon and the ejected proton.
Choosing the $\check{\bf Z}$-direction to be defined
in the direction of $\bf q$, gauge invariance requires
\begin{equation}
\Delta R_L \equiv
\frac{ q_0^2 {\overline{\cal W}}^{00} - q_3^2 {\overline{\cal W}}^{33}  }
     { q_0^2 {\overline{\cal W}}^{00} + q_3^2 {\overline{\cal W}}^{33}  }
\end{equation}
to be zero.
We display results for $\Delta R_L$ as a function of
$Q^2$ for our three canonical nuclei,
$^{12}C$, $^{40}Ca$ and $^{208}Pb$,
 and for both internal oscillator
spacings, $M_2=1.44 \, GeV$ (light)  and $M_2=1.80 \, GeV$ (heavy).

We see, from looking at Figures~\ref{fig:c12drl},
\ref{fig:ca40drl} and~\ref{fig:pb208drl}
that our calculation violates current
conservation at the 10\% level for the DWBA at the lowest energies but only
at the 3-4\% level for the CT cases.  At higher energies, our calculation
becomes more and more current conserving.  Notice that the current is mostly
conserved for the plane wave cases, becoming less so for the heavier
nuclei.

Of course, in nature, current is conserved exactly.  However, we
make the provocative statement that the improvement of current
conservation in our calculation over the traditional Glauber treatment is
more evidence that CT effects are necessary for an accurate theoretical
treatment of $(e,e'p)$ reactions.

\subsubsection{Differential Cross Sections and Normal Polarizations}
\label{sub:dcsnp}

In this section, we display the results from our calculations for the
differential cross sections (as a function of the transverse momentum) and
of the normal polarizations.  These are the quantities which will be measured
in the upcoming CEBAF experiment~\cite{CEBAF,EEF}.  We note that these are
the first predictions of these observables at these energies which include
the effects of CT.  It is important to recall that, in the
CT limit, the normal polarization should vanish.

We only display here, for reasons of space and continuity, some
representative results for
$^{12}C$ nucleus.  The results for the other energies which we calculate
and
for $^{40}Ca$ and $^{208}Pb$ are presented
in Ref.~\cite{GR93}.
Here, the quantity $E_i$ is the initial energy of the incident electron.
Different values of $E_i$ affect only  the kinematical weights used to
construct the cross section and polarization; the nuclear response
functions are independent of the electron kinematics.

Clearly visible, in Figures 14 and 15,
in the shapes of the differential unpolarized cross
sections is the shell structure of the nucleus.  For instance,
$^{12}C$ has 4 $p$-shell protons and only 2 $s$-shell ones.  The
wavefunctions for the $s$-shell nucleons peak at the origin, of course,
while the $p$-shell ones have a node there.  This is the reason why the
cross section for $^{12}C$ has a maximum at about $q_\bot \approx 0.5
\, fm^{-1}$.  Similarly, the shell structure for $^{40}Ca$ and
$^{208}Pb$ can be discerned from the figures in Ref.~\cite{GR93}.

We predict that the
energies proposed in the experiments are {\bf not} high enough to see
the normal polarization vanish.  However,
in $^{12}C$ there
is a measurable suppression of the polarization for the
CT-included case of light excited state mass ($M_2=1.44 \, GeV$).
A detailed comparison with the results of the SLAC NE-18 measurement of
cross section ratios is necessary to see if using the light excited state
mass is still viable. Our preliminary results are that using such a low
mass
may not be inconsistent with the SLAC NE-18 data.
In $^{40}Ca$ and $^{208}Pb$,
there is only a small suppression of the normal
polarization even at $Q^2 = 20 \, GeV^2$.

Thus, it seems that the normal polarization is
a difficult quantity in which to observe CT effects.
But a precise experiment could be
successful.

Note that our formalism does indeed yield vanishing
normal polarization at high energies.  To see this we look
at the observables at
$Q^2 \approx 300 \, GeV^2$,
with optical potential strengths fixed at their
$Q^2 = 5.96 \, GeV^2$ values.
At this extremely high energy, the $EA$
wavefunction is essentially the same as the $OBO_1$ wavefunction and the
normal polarization is seen to be essentially zero for $^{12}C$ over the
range in $q_\bot$ we consider, although for $^{40}Ca$ and $^{208}Pb$ the
polarization is still sizeable at the larger angles, even at this enormous
energy.

In
Figure 16
we show the predictions from the
$DWBA$ and the Born
approximation (BA).  The Born cross section is larger in magnitude
than the $DWBA$ while the polarization is identically
zero for the BA.  The dotdashed curve is the prediction from the exponential
approximation, as before.  The dotted curve is the prediction for the
$LEE_1$ and the dashed curve is for the $OBO_1$.  In all the CT curves,
we have used an excited state mass parameter of
$M_2 = 1.44 \, GeV$.  Much of the differences between the figures for
different electron kinematics is due to the huge difference in the
Mott cross sections, see Table~\ref{tab:kin}.  There is also a sizeable
difference in $V_T$.
However, it is important to
note that all of the
$Q^2 = 5.96 \, GeV^2$ results lead to cross section ratios which are
the same to better than 1\%.
Since absolute magnitudes of the cross sections are larger for the greater
incident electron energies, due to the Mott cross section, these kinematics
are preferred experimentally.
It is also interesting to comment on the
$OBO_1$ curve.  We see that the $OBO_1$ is, already at $Q^2=5.96 \, GeV^2$,
predicting a suppression of the normal polarization.  We know that the
$OBO_1$ is definitely inaccurate at these low energies.  However, the
high-energy nature of the approximation manifests itself in giving small
polarizations.

\subsubsection{Differential Unpolarized and Normal Response Functions}
In this section we display representative graphs of the separated
response functions,
see Figures 17 and 18.
Using these response functions, the total differential
cross sections and normal polarizations can be
constructed by
summing them together with appropriate kinematical weights, given by the
$V$'s.

Since these quantities are more difficult to extract than the total
differential cross sections and polarization, and Saha~\cite{CEBAF,EEF} is
not planning to do so, we only display a few representative figures.
The reader who is interested in more results of this kind are referred to
Ref.~\cite{GR93}.

We see from looking at the response functions that nothing really special
happens for the ``unpolarized response functions''.
Response functions
represent spin-dependent or spin-independent measurements.
It is convenient to adopt the phrases ``unpolarized response''
and ``polarized response'' or ``normal response''
to indicate the spin-independent or spin-dependent (dependent on normal
component) responses.

In lead, there is strong
absorption and the effects of CT do not manifest themselves until very high
energies, making the experimental verification of CT in lead unlikely if
one looks only at unpolarized observables.
However, in the normal responses, we see something quite different and
quite interesting.  In the transverse normal response, $R_T^n$, we see that
for both light and heavy excited state mass cases there is a huge
enhancement in the region of $q_\bot \approx 0.4-1.5 \, fm^{-1}$.  That is,
a ratio of CT response to DWBA response ranges all the way from zero to
infinity over this angular range.

We note that
this CT effect in $R_T^n$ is not of the asymptotic nature one
normally thinks of as CT, lack of absorption, something going to zero,
etc.  However, if one takes the CT wavefunctions we have derived seriously,
then the low energy spin observables are affected in this way.  In this
sense, this enhancement in $R_T^n$ is a pre-asymptotic effect of CT.
This effect is also present in $^{40}Ca$ and in $^{12}C$ but to a
lesser, and still lesser, extent.

If detected, this enhancement in $R_T^n$ would be an unambiguous signature
of CT.  We stress that this enhancement begins at momentum transfers as low
as $Q^2 \approx 2 \, GeV^2$~(!) and continues to the highest energies.
However, the separation of this response function from the total normal
polarization may be difficult.  The kinematical weights, at these energies
and angles, are all approximately equal, see Table~\ref{tab:kin}, so that the
transverse normal response only contributes about one part in twenty to the
polarization.  The predictions presented here indicate that attempting this
separation may be worthwhile.

\subsection{Other Applications}

\subsubsection{Finite Size Effects}
In all of the above calculations we have assumed that the
wavepacket formed in the initial hard collision was of exactly zero size.
This is not realistic.  We do expect, however, that the size of the
wavepacket will decrease with increasing momentum transfer.  It is
interesting to take this non-zero size into account.  This is done by
first considering the form factors defined in
Eqs.~(\ref{eq:thmuct}) and~(\ref{eq:gammamunumn}) to be the product of the
usual form factors with the internal part factored out.  Further, we assume
that the Dirac and Pauli form factors have the same internal dependence.
That is,
we take
\begin{eqnarray}
{F_1}_{N,2m}(q^2) = F_1(q^2) \frac{ f_{N,2m}(q^2) }{f_{N,N}(q^2)},\\
{F_2}_{N,2m}(q^2) = F_2(q^2) \frac{ f_{N,2m}(q^2) }{f_{N,N}(q^2)}.
\end{eqnarray}
In the absence of a compelling reason, note that we choose the internal
dependence for $F_1$ and $F_2$ to be the same.
With this simplification, the internal
form factors simply factor out and (no sum over $\mu$)
\begin{equation}
\left[ \Gamma^\mu (q) \right]^{-1} \Gamma^\mu_{N,2m}(q) =
\frac{ f_{N,2m}(q^2)}{ f_{N,N}(q^2)},
\end{equation}
which is proportional to the unit matrix in Dirac space.
Now, to first order in the interactions, we identify as the
relevant quantities
\begin{eqnarray}
{\cal U}_{cso(so)}^{(2j)f}({\bf B},Z,Z') &=& \sum_{m}
\frac{f_{N,2m}(q^2)}{f_{N,N}(q^2)}
\nonumber\\* &\times &
U_{cso(so)}^{(2m,2j)}({\bf B},Z) e^{i(p_{2m}-p)(Z-Z')},
\end{eqnarray}
where it is understood that, to the present order, only $m=0,1,2$ are
nonzero.  In order to obtain explicit expressions for the internal form
factors, we need to further specify the internal part of the hard
scattering operator.
It is generally believed that the size of the wavepacket is inversely
proportional to the momentum transfer of the virtual photon.  Suppose a
nucleon, comprised of three quarks absorbs a photon of three-momentum
$\bf q$.  Each quark, then, recieves approximately $|{\bf q}|/3$ of the
photon's momentum.  So, we use a form~\cite{BYRON} for the internal part of
the hard operator
\begin{equation}
{\widehat T}_H^\mu(q) = T_H^\mu(q) e^{-{\widehat b} |{\bf q}| /3}.
\end{equation}
This form is suggested by caricatures of perturbative QCD calculations;
see Eq.~(1) of Ref.~\cite{JM91}.

In that case, the internal form factors
$f_{N,N}$, $f_{N,2}$
and $f_{N,4}$ are given by
\begin{eqnarray}
f_{N,N}(x) &=& 1 - x \sqrt{\pi} e^{x^2} \, {\rm erfc}(x),\\
f_{N,2}(x) &=& -x^2 + \left( x^3 + \frac{1}{2} x \right) \sqrt{\pi}
e^{x^2} {\rm erfc}(x),\\
f_{N,4}(x) &=& \frac{1}{2} x^4 + \frac{1}{4} x^2
\nonumber\\* && +
\left( -\frac{1}{2}x^5 - \frac{1}{2}x^3 + \frac{1}{8}x \right)
\sqrt{\pi}e^{x^2} {\rm erfc}(x),
\end{eqnarray}
where $x=x(q^2) = |{\bf q}|b_H / 6$.

The results for non-zero initial wavepacket size are displayed in
Figures 19 and 20.  In
Figure 5 we show the ratio of cross sections for $^{12}C$
as a function of $Q^2$.
As before, the solid line denotes the $DWBA$
(where finite size of wavepacket has no effect)
and CT is included via the exponential approximation for both light
($M_2=1.44 \, GeV$) and
heavy ($M_2=1.80 \, GeV$) excited state masses.  We have also overlayed the
zero size calculations for comparison.  For both the light
(diamonds, dotdashed) at heavy mass (boxes, dashes)
cases, the curves including the finite size are slightly below the zero
size cases.  That the effect is small even for the case of the
light excited state mass
is an
indication that assuming exactly zero size is not a bad approximation.

In Figure 20 we show the CT curves for the current
conservation violation (CCV) quantity $\Delta R_L$ as a function of $Q^2$.
We label the curves $EA$ to indicate that the CT effects are
calculated in the exponential approximation.  The superscripts $LT$ and
$HVY$ indicate the mass of the excited state, light ($LT$) or heavy
($HVY$).  The subscripts $FS$ stand for ``finite size'' and those curves
have included these finite size effects in the manner described above.
Thus, the dotdashed (dashed) curves should be compared with
one another.  Looking at Figure 20
we see that the effects of including finite size improve the
situation as far as gauge invariance is concerned.
Including the size of the initial
wavepacket shows that the CT calculations are gauge invariant at the
2\% level at low energies whereas
the zero size calculations are
current conserving at the 3\% level, at low energies.  However, the trend
to make the calculation better obey the requirements of
current conservation
is encouraging.  We can conclude that assuming the
initial wavepacket to be of exactly zero size introduces only small errors
and can therefore be considered a good approximation.

We do not display results for the finite-sized calculation of
${\cal R}_L / {\cal R}_T$ because there is no discernable difference in
this quantity between the finite-sized curves and the zero-sized ones.

\subsubsection{Fermi Motion}
It has recently been argued~\cite{FSZ93,JK93,JM93,BBK93} that the
Fermi motion of the nucleons bound in the nucleus strongly enhances
the effects of color transparency.  Suppose, in the
$(e,e'p)$ reaction, that the virtual photon three-momentum is labelled by
$\bf q$ and the detected proton three-momentum is called $\bf p$.
Suppose that we call ${\bf p} = {\bf q} + {\bf k}$.  Sometimes $\bf k$ is
called the ``momentum of the struck nucleon''.  This is only true in Born
approximation (the CT limit) because final-state interactions can influence
the outgoing proton's longitudinal momentum.
In exactly quasi-elastic kinematics, the bound nucleon is treated as at
rest and free and so ${\bf k} = 0$.  Fermi motion allows (requires)
${\bf k} \ne 0$.
In the
above calculations, we have assumed that
${\bf k} = {\bf q}_\bot$ where ${\bf q}_\bot$ is very small in magnitude,
$|{\bf p}| \approx |{\bf q}|$, and is purely transverse to $\bf q$.  This
assumption leads to the differential distributions in
Section~\ref{sec:dnr} and in Ref.~\cite{GR93}.

It has been shown~\cite{JK93,JM93,BBK93} that the component of $\bf k$ which is
parallel (or anti-parallel) to $\bf q$ has a huge numerical effect on the
calculated transparencies.  This section is devoted to an investigation of
this effect in our models.

We let the component of $\bf k$ which is in the $\check{\bf q}$ direction
to be called $k_\|$
Thus,
${\bf k} = k_\| \check{\bf Z} + {\bf q}_\bot$.
When ${\bf q}_\bot = 0$ then $k_\| = y$,
where $y$ is the scaling variable introduced in Section~\ref{sub:rlrt}.
Non-zero values of $k_\|$ are especially important in $(p,pp)$
reactions~\cite{FSZ93}  because the experimental results
of Refs.~\cite{CA88,HE89} are presented in terms of bins of $k_\|$.  As
noted earlier, a correct accounting of this effect, in the kinematics of the
Brookhaven experiment, leads to a better understanding of the
data~\cite{JM93}.

We have
calculated the cross sections and other observables for $Q^2 = 0.96$ and
$20.86 \, GeV^2$ for values of $k_\| = -150, -75, 0, 75$ and
$150 \, MeV$.
In particular, we assume that
${\bf p} \cdot \check{\bf Z} = |{\bf q} |+ k_\|$, where $\bf q$ is in the $Z-
$direction.  The results for the
total cross section ratios, ${\cal R}_L / {\cal R}_T$, and the
CCV measure $\Delta R_L$ are shown in
Figure 21.
In all of the figures we display four curves.  The
dotted curve is the Born approximation, the solid curve is the
DWBA, and the dotdashed and the dashed lines are CT results
for light and heavy excited state masses.  In the plots for the cross
section ratios, we display the ratios for the cross section at $k_\|$
divided by the Born cross section at that same value of $k_\|$, except for
the Born curve, which is the cross section at $k_\|$ divided by the Born
cross section at $k_\|=0$.  At low energies, the effect of
finite $k_\|$ is pretty small on the cross section ratios, about
20\%.  At high energies, the effect is quite large for the CT cases,
although still a small effect for the DWBA.  However, nonzero values
for $k_\|$ lead to a large reduction, over a factor of 2,
in the Born cross section at low and
high energies.

Another extremely interesting result from these calculations is shown in
the measure of CCV.  In particular, we see that as $k_\|$ moves away from
zero, current conservation is violated more and more, reaching the
40\% level for $|k_\|| = 150\, MeV$, at low energies.  At higher energies,
the situation improves so that the violations are at the 5\% level.
Clearly, violations of 40\% or more are intolerable and our calculation
probably cannot be trusted there; we shall come back to this point later.
Violations of 5\% probably do not affect the physics too much.  Thus, there
is no problem in trusting our calculation at high energies or at small
values of $k_\|$.

What is the cause of the big change in CCV, due to the non-zero $k_\|$?
One can get a hint by looking at the nuclear current for scalar
electrodynamics.
In that case, the current can be written as
\begin{equation}
J^\mu \propto \int d^4p \, \phi_i (p) \left( p + p' \right)^\mu \phi_f(p'),
\end{equation}
where $\phi_i(p)$ is the momentum space bound state wavefunction and
$\phi_f(p')$ is the scattered wavefunction.  In the Born approximation
where $\phi_f(p') = \delta^4(p'-p-q)$, the integral can be done simply.
The initial nucleon momentum $p$ can now be indentified with $k$, defined
above.  Then, the statement of current conservation is that
\begin{equation}
q_\mu J^\mu \propto  k_\| |{\bf q}|,
\end{equation}
should vanish.  We have used that $k_\mu k^\mu \approx M_N^2$  and chosen
$x=Q^2/(2M_Nq^0)=1$.  Therefore, we see that for vanishing $k_\|$ the
current should be exactly conserved.  Including the spin degrees of
freedom makes this no longer exactly true, but this explains why the Born
approximation is the most current conserving of our models.  This also
explains why, as $k_\|$ gets bigger, the violations of current conservation
 also grow.  For large $Q^2$, we find that $q_\mu J^\mu \propto
Q^2 k_\| / M_N \left[ 1 + {\cal O}(M_N^2/Q2) \right]$,
which explains why the calculations become more gauge invariant at larger
energies.
For non-zero $k_\|$ the above argument indicates that the current is not
conserved.  Why not?
Obviously,
something is missing in our calculation.
The most important cause of CCV in our calculation is probably
coming from the lack of orthogonality
between the initial state and final-state wavefunctions.  Indeed, we
have left out effects in the initial state, such as
particle-hole excitations as well as explicit isobar degrees of freedom
which can alter the nucleon-nucleon force.  Also, we have ignored the
possibility that the nucleons are altered at all in the medium.  Further,
our assumption that the current operator is the free nucleon operator is
probably
not warranted; off-shell effects may be important.  A more careful
treatment of some of these effects could lead to improvements in the gauge
invariance of our models.
The source of these CCV's are of much interest and deserve further study in
the future.

\subsubsection{Lower Components}
We examine the effects of the lower components
of our model wavefunctions here.  We only
present results for the light mass case and for
the $^{12}$C nucleus.

First, we
artificially turn off the lower components in the scattered waves.
The results for the integrated cross section ratios are shown in
Figure 22.
The solid and dotdashed curves are calculated using the
the DWBA and EA wavefunctions with the lower components turned off, divided
by the Born calculation also with no lower components.  The dotted curve in
Figure 22 is the ratio of the Born calculation with no
lower components to the Born calculation with lower components.
The figure shows a striking effect: the lower components are
not significant in the predicted cross section ratios.  These cross section
ratios are seen to be a quantity which is very insensitive to the details
of the wavefunctions.  This is actually quite amazing since the Born
calculation changes so much when the lower components are turned off.

However, we point out that these lower components are crucial
to the approximate
satisfaction of current conservation, see Figure~\ref{fig:c12drllc}.
Further, turning off the lower components has a huge effect on the
integrated ratio of longitudinal to transverse response, see
Figure~\ref{fig:c12rlrtlc}.
Thus, we see that the suppression of the longitudinal response relative to
the transverse response is really entirely located in the lower components.
 This is consistent with the notion, mentioned earlier, that this
longitudinal quenching was due to relativistic effects.  Naturally, since
this ratio is quite different than previously, all of the other observables
such as differential unpolarized cross sections, differential normal
polarizations and differential response functions are also altered by a large
amount.

We have also performed a calculation in which we
take the ratio of lower to upper
components in the scattered wave to be the same as for a free Dirac plane
wave, ${{\bf \sigma}\cdot{\bf p}\over E+M}$.
This replacement has essentially no effect on any of the observables
at the energies we consider.  Only the $\Delta R_L$ is altered noticeably,
with the DWBA violating the current conservation a little more ($\sim
5-10$\%)
without lower
components than with them and the EA conserving the current a little bit
more.  These results show that the CT derivative effects, described in
Section~\ref{sec:de}, are not numerically significant and practically all
of the physics is independent of these CTD's in the lower components.
Since the differences are small we do not display any numerical results
for this calculation.

We also investigate turning off the lower components in the bound state
wavefunction.
This change introduces only very
small corrections to the observables and has only a small
($<$ 5\%) effect on the
level of current conservation violations; the changes in the other
observables are not important.
This is consistent with our expectation; the QHD bound state lower
components contribute only $2-3$\% to the total probability. That is,
$\int dr |F|^2 / \int dr |G|^2 \approx 0.02-0.03$.
 We do not display any
numerical results for this calculation, either.

We conclude that the
lower components are essential to our description of the ejectile
wavefunction.  Without these inherently relativistic components we would
lose approximate current conservation and the experimentally observed
quenching of the longitudinal response, although the integrated cross
section ratios remain unchanged.  However,
examination of the details shows that the
dominant effect in the lower components of the
wavefunctions is the $ \sigma \cdot {\bf p}$ term and
that the complicated terms which appear in the explicit evaluations of the
wavefunctions are unimportant.

We have also explored the effects of changing our bound state
wavefunction.  Instead of using the $QHD$ wavefunction of Horowitz and
Serot~\cite{HS81}, we use harmonic oscillator wavefunctions.
Using these bound state wavefunctions
leads to
only very small changes in the cross section ratios and
in the violations of current conservation.
However, the
differential cross sections are slightly different than those
presented in Section~\ref{sub:dcsnp}.  This is because the $p-$wave
wavefunction has a smaller $\langle R^2 \rangle$ (higher momentum
components) than the corresponding QHD wavefunction.  As a result, the
differential cross section calculated from this oscillator wavefunction
peaks at a larger value of $q_\bot$ and has smaller contribution at low
$q_\bot$.  The contribution to the differential cross section from the
$s-$wave fall faster than the increasing contribution from the $p-$wave,
so that the bump that we saw in the QHD case disappears.

It is possible to fix the $\langle R^2 \rangle$ for the harmonic oscillator
wavefunctions to agree with that predicted by the QHD model.  Doing this
improves the agreement between the $p-$waves but at the expense of the
agreement between the $s-$waves.  Thus, the differential cross sections
seem to be sensitive to the precise form of the bound state wavefunctions.

However, the main conclusion of this work on polarization, that the normal
polarization is not hugely suppressed at present energies,
remains unchanged when harmonic oscillator bound
states are used.

\section{Summary and Conclusions}
\label{sec:end}
The effects of color transparency
(CT) in quasi-elastic
$(e,e'p)$ reactions at large momentum transfers are explored here.
We have focused on the $(e,e'p)$ reaction since lepton scattering is simpler
than
hadronic scattering.  Of course, CT effects should also be present in other
reactions such as $(p,pp)$, $(\pi, \pi p)$, etc.  The methods presented in
this paper can be applied to these reactions.

In this paper we include
the effects of proton and photon spin.  Thus, we call this work
``Vector CT'', reflecting the vector nature of the photon.  So far no
published work has included the effects of spin in calculations of CT.
The main motivation for undertaking this calculation is the proposal by
Saha and collaborators~\cite{CEBAF,EEF} who plan to measure the normal
component of the proton polarization in $(e,e' \vec p)$ experiments.
This
polarization is an interesting quantity because it
vanishes in the absence of final-state interactions.
Further, the photon really is a vector particle and its spin should play
some role in the scattering.
Dirac
phenomenology is used to construct the DWBA.
The internal
operators and states are then embedded
in to this formalism.
We obtain detailed expressions for the
CT wavefunction in terms of these internal operators.

In order to calculate the CT effects, it is necessary in this
approach to choose an explicit
form for the wavepacket-nucleon interaction.  Although
this interaction is not precisely known, some general properties are
generally agreed upon.  We choose a simple representative form for this
interaction which is consistent with the known constraints.

Further, it is also necessary
to assume a model for the internal space on which the CT operators act.
For simplicity, we choose the baryon spectrum to be represented by a
two-dimensional transverse harmonic oscillator.  With this choice comes a
single free parameter, the oscillator spacing, which determines the masses
of the nucleon resonances.  This spacing is characterized by the mass
of the first even parity excited state, which we label by $M_{N^*}$.  In
this paper, we have chosen the values $M_{N^*}=1.44 \, GeV$
and $M_{N^*} = 1.80 \, GeV$ for the
numerical results we display.  Increasing this mass
postpones the onset of CT to higher energies.

We evaluate the formal distorted wavefunction by making various
approximations to sum up different parts of the multiple-scattering series.
In each energy regime we define an unperturbed piece of the path-ordered
exponential, which we can solve exactly,
and a perturbed piece, which we treat systematically in
perturbation theory.  We have developed high-energy and low-energy
approximations.  However, the different approximation schemes overlap
so that we are confident that we can accurately
approximate the exact distorted wavefunction at all energies
(above 1 $GeV$).  In particular, we find that simply ignoring the path
ordering symbol in the wavefunction serves as a good approximation of
the full distorted wavefunction for all energies.

Although we use over-simplified models for the wavepacket-nucleon
interaction and the internal baryon space, the methods of calculation
presented in this paper are more general and can be used with more
realistic interactions and models.

We calculate ratio of cross sections for $^{12}C$, $^{40}Ca$ and
$^{208}Pb$ as a function of $Q^2$.  In contrast to earlier calculations
which assume a constant value of $40 \, mb$ for the elementary
proton-proton cross section, we take the cross section and other optical
potential strengths directly from data.  Thus, we confirm
the assertion of Frankfurt, Strikman and Zhalov, that energy dependence
in $\sigma/\sigma^B$ is expected.
This is because the $pp$ cross section, in
the energies of the experiment varies in such a way that the Glauber
treatment decreases the transparency ratio.  Therefore, since the
preliminary results of the
experiment, so far, see a small variation
in the ratio as a function of $Q^2$, there may
be some transparency effects to compensate.  We conclude that the
excited state mass which is consistent with the preliminary data is about
$M_{N^*} \approx 1.6 \, GeV$ or more.

A typical problem with calculations of this sort is the lack of current
conservation.
We show that the violations of gauge invariance are at the 10\% level
for the DWBA and
only at the 4\% level for the CT wavefunctions at low energies but are
all at the $1-2$\% level at the highest energies, $Q^2 \approx 20 \,
GeV^2$.  Inclusion of CT effects is therefore desirable, even at low
energies, because it improves the situation from the DWBA with respect to
current conservation and gauge invariance.

Because of the extra spin degrees of freedom, there are many many other
quantities which we calculate.
We display results for the ratio of
integrated longitudinal to transverse response as well as for differential
cross sections and normal polarizations.  The differential
cross sections do display a noticeable, and detectable, increase at large
momentum transfers.  This is just the data which is summarized in the
integrated cross section ratios described above.  In the normal
polarizations, we see that the deviations from the usual Glauber treatment
do not seem to be significant or measurable at any energy.
Although $Q^2 \approx 20 \, GeV^2$ we see, in $^{12}C$ a moderate
decrease of the
polarization for the light mass CT case but not for the heavy case.
We conclude that the energies at
which the normal polarization should completely vanish are, unfortunately,
quite high. Frankfurt,
Strikman and Zhalov have obtained a similar result (private communication).

We have also calculated
the individual
separated response functions, eight in all for in-plane scattering, which
require many measurements at many angles in order to accurately extract
from the cross sections and polarizations.
We notice, in some of the response functions, that
some pre-asymptotic behavior of the CT wavefunctions are manifest are
possible detectable.  This is an effect which increases in heavy nuclei
like $^{208}Pb$.  In particular, if it is possible to separate out from the
normal polarization the responses $R_L^n$, $R_T^n$, $R_{TT}^n$ and
$R_{LT}^n$, then it should be possible to see in $R_T^n$ a {\em huge}
enhancement
over the Glauber result at $|{\bf q}_\bot| \approx 0.5-1 \, fm^{-1}$ for
$Q^2 \ge 2 \, GeV^2$!.
This
extraction may be difficult because, although the kinematical weights are
all approximately equal, $R_T^n$ contributes only about one part in
ten at most to the total polarization.

Further, we have also investigated the effects of non-zero wavepacket size
and of Fermi motion.  Including the non-zero size of the initial wavepacket
produces only small changes so its neglect is a very good approximation.
This is because of effects of wave packet expansion.  Very small sized wave
packets expand quickly into small sized wave packets.
However, a slight improvement in the current conservation is obtained by
including the non-zero size.
The Fermi motion, on the other hand, can have a big
effect on the cross section ratios if the ``initial nucleon momentum'' is
along the direction of $\bf q$.  However, for values of
$k_\|$ ${\mathrel{\rlap{\lower4pt\hbox{\hskip1pt$\sim$}}
\raise1pt\hbox{$>$}}}$ 150 MeV/c,
which is this component, the violations of
current conservation
become severe at low energies and our calculation cannot be trusted there.
At high energies, the calculation becomes more gauge invariant and we see
an interesting effect for non-zero $k_\|$.  This effect is especially
important for the $(p,pp)$ reaction where the data is displayed in bins of
$k_\|$~\cite{JM93}.

Lastly, we have also investigated the effects of the lower
components on cross section ratios, current conservation, polarization,
etc.  The complicated nature of the DWBA and CT lower
components are not significant
and taking the ratio of lower to upper
components for these distorted waves to be the same as for plane waves
introduces no discernable results
(CTD effects are unimportant).  The bound state lower components are
similarly seen to be unimportant in the prediction of most of the
observables.  Finally, we have investigated the effects of using harmonic
oscillator bound states instead of the ones given by QHD.  The cross
section ratios are essentially the same as for the QHD bound states,
while the differential cross sections, polarizations and response
functions all differ somewhat from those using QHD bound states.

In conclusion, we can make the following
short comments.  The recent SLAC data~\cite{McK92} imposes constraints on
the allowable models shown in this paper.  Despite the lack of significant
$Q^2$
variation in the data it is possible, because of the energy dependence of
the elementary $pp$ observables, that the data actually may be an
example of the manifestation of CT.  A measurement of the normal
polarization in $(e,e' \vec p)$ reactions does not seem to be a good way to
see CT effects at moderate $Q^2$.  However, a measurement of the normal
transverse response in a heavy nucleus such as $^{208}Pb$ does seem to
afford the opportunity to see CT, unambiguously,
at quite low momentum transfers.

\acknowledgments{We wish to thank B. Jennings, L. Frankfurt and M.
Strikman for useful discussions, and C. Horowitz for his computer
code.  This work
is supported in part by the U.S. Department of Energy.}

\appendix
\section{Extraction of Optical Potential Strengths}
\label{app:nnamp}

We describe how we obtain the strengths of the scalar
and vector optical potentials, first shown in Eq.~(\ref{eq:vsvv}), and
summarized numerically in Table~\ref{tab:strengths}.  In particular, we
extract these strengths from proton-proton elastic scattering data.  In
order to show exactly what we have done, it is useful to recall some
aspects of nucleon-nucleon scattering.

\subsection{Nucleon-Nucleon Scattering}

We use the par\-ameter\-ization of the nucleon-nucleon scat\-ter\-ing
amp\-li\-tude
of~\cite{SMW83,MRW83,MSW83,WA81,WA84,WA86,WA87},
\begin{eqnarray}
\frac{f_c}{2ik} &=& A + B \mbox{\boldmath $\sigma $}_1 \cdot
\mbox{\boldmath $ \sigma $}_2 +
iq C \left( \sigma_{1n} + \sigma_{2n} \right)
\nonumber\\* && \qquad \qquad + D
\mbox{\boldmath $ \sigma$}_1
\cdot {\bf q} \,
\mbox{\boldmath $ \sigma$}_2 \cdot {\bf q} + E \sigma_{1Z} \sigma_{2Z}.
\label{eq:amp2}
\end{eqnarray}
The amplitudes $A, B, C, D, E$ can be easily related to the s-channel
helicity amplitudes~\cite{GGMW60}.

One can also parameterize the NN scattering amplitude in a Lorentz
invariant form~\cite{SMW83,MRW83,MSW83,WA81,WA84,WA86,WA87},
by describing the spin
dependence in terms of Dirac matrices,
\begin{eqnarray}
F &=& F_s + F_v \gamma_1^\mu \gamma_{2 \mu} + F_t \sigma_1^{\mu \nu}
\sigma_{2 \mu \nu} + F_p \gamma_1^5 \gamma_2^5
\nonumber\\* && \qquad \qquad + F_a \gamma_1^5
\gamma_2^5 \gamma_1^\mu \gamma_{2 \mu}.
\end{eqnarray}
Naturally, the quantities $F_s$, $F_v$, etc. are functions of the momentum
transfer, $\bf q$.
By taking the matrix element of this operator and doing the Dirac algebra,
leaving $F$ to be an operator in the Pauli spin space, we can equate the
two descriptions, modulo some kinematical factors.  These factors are
known and are derivable from the relativistic impulse
approximation~\cite{SMW83,MRW83,MSW83,WA81,WA84,WA86,WA87}.
In fact, the relationship
is, in the center-of-mass frame,
\begin{eqnarray}
\chi_{\lambda_1'}^\dagger
\chi_{\lambda_2'}^\dagger \, \frac{f_c}{2ik} \,
\chi_{\lambda_1}\chi_{\lambda_2},
&=& \nonumber\\* &&
\! \! \! \! \! \! \! \! \! \! \! \!
\! \! \! \! \! \! \! \! \! \! \! \!
\! \! \! \! \! \! \! \! \! \! \! \!
\! \! \! \! \! \! \! \! \! \! \! \!
\overline{u}({\bf k'}, \lambda_1')
\overline{u}(-{\bf k'}, \lambda_2') \, F \, u({\bf k}, \lambda_1)
u(-{\bf k}, \lambda_2)
\end{eqnarray}
where $\chi_\lambda$ are Pauli spinors for helicity $\lambda$ and
$u({\bf k}, \lambda)$ are free Dirac spinors corresponding to momentum
${\bf k}$ and helicity $\lambda$; $k$ is the center-of-mass momentum.

\subsection{The Lorentz Invariant Forward Amplitudes}
We are concerned with the case of elastic scattering of protons from nuclei.
It is in this case that we can extract the optical potential strengths.
Within the context of the relativistic impulse approximation, the T-matrix
for nucleon-nucleus scattering is simply a sum of the t-matrices of the
elementary nucleon-nucleon scattering.  If we neglect nuclear-medium
modification to the bound nucleons, off-shell effects, and $1/A$
corrections, the optical potential is approximately equal to the
impulse approximation T-matrix~\cite{MSW83}.
In a spin saturated nucleus, the matrix element of this Dirac operator
involves a trace over struck nucleon spins which eliminates all terms
except the tensor, scalar and the time component of the vector terms.
Thus, in
this relativistic impulse approximation,
the optical potential is of the form,
\begin{eqnarray}
U_{opt}({\bf q}) &=& -\frac{4\pi i p_{lab}}{M_N} \Bigl[F_s(q) \rho_s(q)
+ \gamma_1^0 F_v(q) \rho_v(q)
\nonumber\\* && \qquad \qquad -
\frac{ \mbox{\boldmath $\alpha$} \cdot
{\bf q}}{M_N} F_t(q) \rho_t(q) \Bigr],
\end{eqnarray}
where $\rho_s$, $\rho_v$ and $\rho_t$
 are scalar, vector and tensor nuclear densities and
$p_{lab}$ is the lab frame momentum.
The kinematical factors on the right
hand side of the above equation,
$r \equiv -4 \pi i p_{lab} / M_N$,
 come from the relationship between the
Lorentz invariant Feynman amplitude and the usual description of the
optical potential in the impulse approximation.

The coordinate space optical potential is obtained by Fourier transforming
the momentum space potential.
In the limit of very high energies, the scattering is predominantly in the
forward direction.  Thus, we can approximate $F_s(q)$ and $F_v(q)$ by their
forward direction values $F_s^0$ and $F_v^0$, and neglect the tensor term
since $q$ is small.
Thus, we arrive
at~\cite{SMW83,MSW83}
\begin{equation}
U_{opt}({\bf R})=   r\, \left[F_s^0 \rho_s(R)
+ \gamma_1^0 F_v^0 \rho_v(R) \right].
\end{equation}

In the forward scattering approximation, the relationship between the Pauli
and Dirac amplitudes can be written as a matrix equation~\cite{MRW83}.
Since $A,B,C,D,E$ are to be taken directly from data,
we can simply invert this matrix to obtain the scalar and vector
density strengths from the Pauli amplitudes.  In particular, we find
\begin{eqnarray}
F_s^0 &=& \frac{1}{2\xi^2(1+\xi)} \biggl[ \xi (2\xi+1) A^0
+ (2\xi^2-1) B^0
\nonumber\\*
&&
+ 2M_N \xi (2 \xi^2-1) \sqrt{ \frac{\xi+1}{\xi-1} }C^0
+ \frac{1-2 \xi^2}{2( \xi-1 ) } E^0 \biggr],
\label{eq:fs0}\\
F_v^0 &=& \frac{1}{2 \xi^2(1+ \xi)} \biggl[ \xi A^0 - B^0
-2M_N \xi \sqrt{ \frac{ \xi + 1}{\xi-1} } C^0
\nonumber\\* && \qquad \qquad
\qquad \qquad \qquad + \frac{1}{2( \xi-1 ) } E^0
\biggr].
\label{eq:fv0}
\end{eqnarray}
where $\xi=\sqrt{s}/2M_N$ is dimensionless
Notice that the quantity $D^0$ does not enter in for the scalar and vector
strengths.
It is interesting to realize that inasmuch as the central potential
dominates, it is really only the quantities $A^0$ and
$C^0$ which are important.
Indeed, if the differences between the scalar and vector nuclear
densities are neglected, a relevant quantity which appears in the
Dirac eikonal formalism is
\begin{equation}
- \frac{2M_N r F_s^0 + 2 E_{lab} r F_v^0}{2ip_{lab}} = 4 \pi A^0,
\end{equation}
That this should be true is not a surprise.
$A^0$ is just the forward scattering
amplitude, whose imaginary part is proportional to the total cross section.

\subsection{Extracting Forward Amplitudes from Data}
The problem is how to extract the NN parameters
$A^0, B^0, C^0, E^0$ from the data.  We do this in two ways.  In the first
way, we take some recent phase shift solutions to reconstruct the helicity
amplitudes and then to use the above relations to get at the
quantities we need.
One must be very careful,
however, since there exist many different parameterizations of the
experimental phase shifts and of the scattering amplitude so that one must
make sure that all of the parameterizations are consistent with one another.
The second method involves a more direct determination of the relevant
amplitudes from data.

\subsubsection{Forward Amplitudes from $Q^2 = 1 - 6 \, GeV^2$}
Wallace has published~\cite{WA81}
the Pauli amplitudes,
$A^0,B^0,C^0,D^0,E^0$
for Hoshizaki's phase shift solutions up to $Q^2 = 4 \, GeV^2$.
In particular, Wallace uses~\cite{SJW} a more recent solution~\cite{HO78}
than that published in
Refs.~\cite{HK69,HK76,HO76}, as well as calculating at more energies.
Therefore, we construct the optical potential
strengths in this energy regime by using Eqs.~(\ref{eq:fs0})
and~(\ref{eq:fv0}) for Wallace's amplitudes.

At higher energies, we use
the phase shift analysis
of Higuchi and Hoshizaki~\cite{HH79} for $pp$ phase shifts at
$p_{lab} = 4 \, GeV$, which translates to $Q^2= 5.96 \, GeV^2$, in order to
reconstruct the T-matrix and extract the strengths as outlined.
Using this method, we were able to construct the T-matrix
and to reproduce the experimental data and theoretical
predictions published by
Hoshizaki~\cite{HH79}.
This serves as a consistency check that we are using the phase shifts
correctly.

We must stress the fact that at these energies the phase shifts are only
loosely constrained, at best.  Therefore, different phase shift analyses
will  yield different values for the optical potential strengths.  However,
our strengths are roughly
consistent with those found in Ref.~\cite{MRW83}.

\subsubsection{Forward Amplitudes at $Q^2 = 9.65$ and $20.86 \, GeV^2$}
Phase shift solutions provide a powerful method to summarize the scattering
data and to predict observables at all momentum transfers.  At higher
energies, more and more partial waves become important (and the data
becomes more and more scarce) and phase shift analyses become difficult.
However, phase shift analyses do exist~\cite{MSWY81,MSY86} for (lab)
beam momenta of 6 $GeV/c$ and 12 $GeV/c$; in quasi-elastic kinematics,
these beam momenta translate to $Q^2 = 9.65$ and $20.86 \, GeV^2$.
However, the lack of
constraining data as well as the idea that the high angular momentum
partial waves are not negligible and must be modelled leads us away
from the phase shifts.  We will, however, take certain qualitative features
from these analyses; see below.
Further, the phase shift analyses do not emphasize the forward direction
observables over the other angles.  Indeed,
\begin{eqnarray}
A^0 &=& \frac{\sigma}{8\pi} \left( 1 - i \alpha_f \right),\\
B^0 &=& \frac{\Delta \sigma_T}{-16 \pi} - i \frac{F_2}{4k},\\
E^0 &=& \frac{\Delta \sigma_L}{-16 \pi} + i \frac{F_3}{4k} - B^0,
\end{eqnarray}
where $\alpha_f$ is the ratio of the real part to the imaginary part of the
forward scattering amplitude and $k$ is the center-of-mass momentum.
The other real parts are
$F_2 = \frac{p_{lab}}{k}{\rm Re \,}\phi_2(0^\circ)$ and
$F_3 = \frac{p_{lab}}{k}{\rm Re\, } \left[\phi_1(0^\circ) - \phi_3(0^\circ)
\right]$.  The longitudinal cross section, $\Delta \sigma_L$, is given by
the difference of measured cross sections with beam and target proton spins
aligned and anti-aligned with the beam direction
with that of the both spins aligned along the beam direction.
Similarly, the transverse
cross section is measured by subtracting cross sections where the beam
proton spin is aligned with the normal to the beam direction and the target
spin is anti-aligned minus the cross section where both beam and target
spins are aligned with the normal to the beam direction.
The quantities $F_2$ and $F_3$ are evaluated using dispersion
relations~\cite{GK78}.

Thus, the only required quantity not available directly from
the forward cross section data is the parameter $C^0$.  However, this
parameter can be obtained if enough data on the spin observables at small
$-t$ is known.  For instance, in the four component notation of
Refs.~\cite{BLW78,BLL87,LA90}, we can write two of the spin observables as
\begin{eqnarray}
\frac{k^2}{\pi}\frac{d\sigma}{dt} \, P_n & = &
 {\rm Re \,} a^*e,\\
\frac{k^2}{\pi}\frac{d\sigma}{dt} \,
A_{00sk} &=& -{\rm Re \,
}a^*d \sin \theta - {\rm Im \, }d^*e \cos \theta,
\end{eqnarray}
where $\theta$ is the center-of-mass scattering angle and
\begin{eqnarray}
A &=& \frac{a+b}{4ik},\\
B &=& \frac{a-b}{4ik},\\
C &=& -\frac{e}{4kq},\\
D &=& \frac{-a+b+c-d}{4ik},\\
E &=& \frac{-a+b+c+d}{4ikq^2}.
\end{eqnarray}
Notice that both of
these quantities, $P_n$ and $A_{sk}$ ($A_{sk}\equiv A_{00sk}$),
vanish in the forward direction ($\theta=e=0$). Their slopes (as a function
of $q$, not $-t$) at
$q=0$, however, are finite.  Taking the derivatives with respect to
$q=|{\bf q}| = \sqrt{-t}$ and evaluating in the forward direction, we obtain
\begin{eqnarray}
\frac{k^2}{\pi} \left( \frac{d\sigma}{dt} \right)_{t=0}
\left( \frac{dP_n}{dq} \right)_{q=0} &=& {\rm Re \, }a^* \frac{e}{q},\\
\frac{k^2}{\pi} \left( \frac{d\sigma}{dt} \right)_{t=0}
\left( \frac{dA_{sk}}{dq} \right)_{q=0} &=& -\frac{1}{k} {\rm Re \,} a^*d
- {\rm Im \,} d^* \frac{e}{q},
\end{eqnarray}
where we have used $e/q \approx e'(q)$ in the forward
direction.  Note that $e/q$ remains finite in this limit.
We have also changed variables and written the differential
cross section in Lorentz invariant form.  Now, denoting $x_R$ and $x_I$ as
the real and imaginary parts of the complex number $x$,
we can use the above two equations to write
\begin{eqnarray}
C_R &=& -\frac{a_I}{4k^2} - \frac{k}{4\pi} \frac{1}{a_R d_R + a_I d_I}
\left( \frac{d\sigma}{dt} \right)_{t=0} \times \nonumber\\*
&& \qquad \left[
d_R \left( \frac{dP_n}{dq} \right)_{q=0} + a_I \left( \frac{dA_{sk}}{dq}
\right)_{q=0} \right],\\
C_I &=& -\frac{a_R}{a_I}C_R - \frac{k}{4\pi a_I}
\left( \frac{d\sigma}{dt} \right)_{t=0} \left( \frac{dP_n}{dq} \right)_{q=
0},
\end{eqnarray}
where a superscript $^0$, indicating the forward direction,
is to be understood on the quantities $C_R, C_I, a_R, a_I, d_R, d_I$.

Thus, with these equations, and some good spin data, we can extract the
optical potential strengths directly from data.  We note that we have
tested these equations for the spin parameter $C$ on the 4 $GeV/c$ data of
Hoshizaki~\cite{HH79} and they agree with the value calculated strictly by
using phase shifts.  This gives us confidence that these equations are
reliable to use at higher energies.

The experimental values for the $\sigma^{tot}$, $\alpha_f$,
$\Delta \sigma_L$, $\Delta \sigma_T$, $F_2$ and $F_3$ are sum\-mar\-ized in
Refs.~\cite{MSWY81,MSY86}.  In order to complete the analysis described
above, we need information on the slopes of $P_n$ and $A_{sk}$ at small
$q$.  In Refs.~\cite{RU75,KL77} the polarization $P_n$ has been measured
at $p_{lab}=6\, GeV/c$.  Better than that, an empirical fit to
the data is given, which can be directly differentiated.  By averaging these
two
values from the two references, we obtain the result
\begin{equation}
\left( \frac{dP_n}{dq} \right)_{q=0} = 0.494 \, GeV^{-1}
\quad {\rm for \quad p_{lab} = 6 \,}GeV/c.
\end{equation}
The uncertainty in this quantity is only about 3\%; see
Refs.~\cite{RU75,KL77}.
At $p_{lab}=12\, GeV/c$, the polarization has been
measured~\cite{KR78} and again an empirical fit is given.  Thus,
differentiating this quantity directly we find that
\begin{equation}
\left( \frac{dP_n}{dq} \right)_{q=0} = 0.215 \, GeV^{-1}
\quad {\rm for \quad p_{lab} = 12 \,}GeV/c.
\end{equation}

To proceed, we now notice something amazing.  Looking at the data for
$A_{sk}$ at $p_{lab}=4 \, GeV/c$~\cite{HH79} we see that the
experimental slope of
$A_{sk}$ is zero to a very good approximation.  Indeed, explicit
calculation based on Hoshizaki's phase shifts yield a slope of
0.014 $GeV^{-1}$.  There is very little data for $A_{sk}$ at higher
energies, yet the phase shift solutions~\cite{MSWY81,MSY86} also suggest
that the slope of $A_{sk}$ is very small.  Based on this evidence, we
assume that
\begin{equation}
\left( \frac{dA_{sk}}{dq} \right)_{q=0} = 0 ,
\end{equation}
at least for energies below $p_{lab}=12 \, GeV/c$.

Using the central values in the published data yields optical potential
strengths which are larger in magnitude than the lower energy values.  However,
by adjusting
the parameters within their experimental error bars,
numbers in better agreement (more consistent) with the low energy data can be
obtained.
It is these numbers which are summarized in
Table~\ref{tab:strengths}.  There is therefore considerable uncertainty in
the last two rows of Table~\ref{tab:strengths}.
In particular, the forward scattering data
we use to generate the strengths are displayed in Table~\ref{tab:fsdata}.
The intermediate Pauli amplitudes obtained in the manner described in this
appendix are displayed in Table~\ref{tab:pauliamps}.

Clearly, the most important of these parameters are $A^0$, which contains
the information about the forward scattering amplitude, and the spin-flip
parameter $C^0$ which is determined from the slopes of the forward spin
observables.

\section{DWBA and CT Eikonal Wavefunctions}

\label{app:vectorctwf}

\subsection{DWBA Eikonal Wavefunction}

The expression for the DWBA wavefunction,
Eq.~(\ref{eq:ufam}), can be simplified by
performing the spin algebra.  First, we note that in the coordinate system
$(B,Z,\phi)$, the cross product
${\bf B} \times {\bf p} = -Bp \check{\mbox{\boldmath $\phi$}}$
($B = |{\bf B}|$ and $p = |{\bf p}|$).
To reduce the wavefunction to its fundamental dependence (linear) on the
spin matrices, we define the quantities
\begin{eqnarray}
\omega_1^{(+)}({\bf B},Z) &=& \int_{-\infty}^Z \frac{dZ'}{2ip}
\biggl[ U_c({\bf B},Z')
\nonumber\\* && \qquad - ipZ' \, U_{so}({\bf B},Z') \biggr],
\label{eq:omega1+}\\
\omega_2^{(+)}({\bf B},Z) & = &
\frac{B}{2}\int_{-\infty}^Z dZ' \, U_{so}({\bf B},Z').
\label{eq:omega2+}
\end{eqnarray}
Then, we obtain
the final DWBA wavefunction as
\begin{equation}
\Psi^{(+)}_{{\bf p}, -\check{\bf s}_R}({\bf R}) =
{\cal N} \left( \begin{array}{cc}
u_{\bf p}^{(+)} ({\bf R}) \\ w_{\bf p}^{(+)} ({\bf R}) \end{array} \right)
\, \chi_{-\check{\bf s}_R},
\end{equation}
where
\widetext
\begin{eqnarray}
\lefteqn{
u_{{\bf p}}^{(+)} = e^{ipZ}e^{\omega_1^{(+)}}
\left[ \cos \omega_2^{(+)} + i \mbox{\boldmath $\sigma$}
\cdot \check{\mbox{ \boldmath $\phi$ }} \sin
\omega_2^{(+)} \right] },
\label{eq:udwba}\\
\lefteqn{
w_{{\bf p}}^{(+)} = e^{ipZ}e^{\omega_1^{(+)}}
\frac{-i}{E+M_N+V_s-V_v} \qquad \times} \nonumber\\*
&&
\Biggl\{ \mbox{\boldmath $\sigma$} \cdot {\check{\bf Z}}
\left[ \cos\omega_2^{(+)} \left( \omega_{1,Z}^{(+)}
-\omega_{2,B}^{(+)}+ip \right) -\sin\omega_2^{(+)} \left(\omega_{1,B}^{(+)}
+\omega_{2,Z}^{(+)}+\frac{1}{B} \right) \right] \nonumber\\*
&&+\mbox{\boldmath $\sigma$} \cdot
{\check{\bf B}} \left[ \sin\omega_2^{(+)} \left( \omega_{1,Z}^{(+)}
-\omega_{2,B}^{(+)}+ip \right) +\cos\omega_2^{(+)} \left(\omega_{1,B}^{(+)}
+\omega_{2,Z}^{(+)}\right) \right] \Biggr\},
\label{eq:wdwba}
\end{eqnarray}
\narrowtext
where
\begin{eqnarray}
\omega_{i,Z}&=&\frac{\partial}{\partial Z} \omega_i,\\
\omega_{i,B}&=&\frac{\partial}{\partial B} \omega_i,
\end{eqnarray}
and where we have suppressed the argument $\bf R$ in $\omega_1$,
$\omega_2$, $V_s$ and $V_v$.

We note that
the wavefunction for incoming boundary conditions is obtained simply by
taking the complex conjugate of the potentials and changing the limits of
integration.  Thus,  $\Psi_{{\bf p},-\check{\bf s}_R}^{(-)}$
has the same form as
Eqs.~(\ref{eq:udwba}) and~(\ref{eq:wdwba}) except that
$\omega_i^{(+)} \rightarrow \omega_i^{(-)}$, where $i=1,2$ and
\begin{equation}
\omega_1^{(-)}({\bf B},Z)
= - \int_Z^{\infty} \frac{dZ'}{2ip}
\left[ U_c^*({\bf B},Z') + ipZ' \, U_{so}^*({\bf B},Z') \right],
\label{eq:omega1-}
\end{equation}
\begin{equation}
\omega_2^{(-)}({\bf B},Z)
 =  -\frac{B}{2}\int_Z^{\infty} dZ' U_{so}^*({\bf B},Z'),
\label{eq:omega2-}
\end{equation}
where $U^*$ represents the complex conjugate of the quantity $U$.

It is also instructive to consider the case where the optical potentials
are small, in some sense.  Then, we can perform an order by order
expansion of the DWBA wavefunction, in terms of the potentials.  This is
useful as a practical matter because the size of the first-order term of
the DWBA is a good indicator of how accurate the $OBO$ CT wavefunction will
be, see
Ref.~\cite{GM93}.
Thus, we proceed by expanding in powers of $\omega_1$ and neglecting
terms of order
${\cal O}(\omega_i^3)$.  We obtain,
\begin{equation}
\Psi^{(+)}_{{\bf p}, -\check{\bf s}_R}({\bf R}) = e^{ipZ} \sum_m
\Psi_{m, {\bf p}, -\check{\bf s}_R}^{(+)}({\bf R}),
\end{equation}
where the subscript $m$ indicates the order of the expansion of the
potentials.  Decomposing the wavefunction in terms of upper and lower
components,
\begin{equation}
\Psi_{m, {\bf p}, -\check{\bf s}_R}^{(+)}({\bf R}) = {\cal N} \left(
\begin{array}{c}
u_{m, {\bf p}}^{(+)} ({\bf R}) \\ w_{m, {\bf p}}^{(+)}({\bf R}) \end{array}
\right) \chi_{-\check{\bf s}_R}.
\end{equation}
We present explicit expressions for the order
by order expansion of the DWBA wavefunction.  We only need the first
few terms.
\widetext
\begin{eqnarray}
u_{0,{\bf p}}^{(+)} &=& 1,
\label{eq:udwba0}\\
w_{0,{\bf p}}^{(+)} &=&
\frac{ \mbox{\boldmath $\sigma$} \cdot {\bf p}}{E+M_N+V_s-V_v} ,
\label{eq:wdwba0}\\
&&\nonumber \\
u_{1,{\bf p}}^{(+)} &=& \left( \omega_1^{(+)} + i\omega_2^{(+)}
\mbox{\boldmath $\sigma$} \cdot \check{\mbox{\boldmath $\phi$}} \right),
\label{eq:udwba1}\\
w_{1,{\bf p}}^{(+)} &=& \frac{1}{E+M_N+V_s-V_v}
\nonumber\\* && \qquad \times
\Biggl\{ \mbox{\boldmath $ \sigma$} \cdot
{\check{\bf Z}} \left[ p\omega_1^{(+)} -i \omega_{1,Z}^{(+)}
+i \omega_{2,B}^{(+)} +i \frac{\omega_2^{(+)}}{B}
 \right] \nonumber\\*
&&\qquad
+\mbox{\boldmath $\sigma$} \cdot {\check{\bf B}} \left[ p\omega_2^{(+)}
- i \omega_{1,B}^{(+)}
- i \omega_{2,Z}^{(+)}\right] \Biggr\},
\label{eq:wdwba1}\\
&&\nonumber \\
u_{2,{\bf p}}^{(+)} &=& \left[ \frac{1}{2} \left( \omega_1^{(+)}
\right)^2 - \frac{1}{2} \left( \omega_2^{(+)} \right)^2 + i
\mbox{\boldmath $\sigma$} \cdot \check{\mbox{\boldmath $\phi$}} \,
\omega_1^{(+)} \omega_2^{(+)} \right] ,
\label{eq:udwba2} \\
w_{2,{\bf p}}^{(+)} &=&  \frac{1}{E+M_N+V_s-V_v}
\Biggl\{ \mbox{\boldmath $\sigma$} \cdot {\check{\bf Z}} \Bigl[ \frac{p}{2}
\left( \omega_1^{(+)} \right)^2
-\frac{p}{2} \left( \omega_2^{(+)} \right)^2 \nonumber \\*
&& -i \omega_1^{(+)} \omega_{1,Z}^{(+)} + i \omega_2^{(+)}
\omega_{2,Z}^{(+)} + i \omega_1^{(+)} \omega_{2,B}^{(+)}
+i \omega_2^{(+)} \omega_{1,B}^{(+)}
+ i \frac{ \omega_1^{(+)} \omega_2^{(+)} }{B} \Bigr] \nonumber \\*
&& + \mbox{\boldmath $\sigma$} \cdot {\check{\bf B}}
\biggl[ p \, \omega_1^{(+)} \omega_2^{(+)}
-i \omega_1^{(+)} \omega_{1,B}^{(+)} + i \omega_2^{(+)} \omega_{2,B}^{(+)}
\nonumber \\* &&
-i \omega_2^{(+)} \omega_{1,Z}^{(+)} - i \omega_1^{(+)} \omega_{2,Z}^{(+)}
\biggr] \Biggr\}.
\label{eq:wdwba2}
\end{eqnarray}
\narrowtext

\subsection{CT Eikonal Wavefunctions}
Here, we show the explicit CT eikonal wavefunctions in our
various approximation schemes.
More notation is needed, so we
introduce the quantity
\begin{equation}
{\cal U}_{cso}^{(2j)} ({\bf B},Z,Z') = {\cal U}_c^{(2j)}({\bf B},Z,Z')
- i p Z' \, {\cal U}_{so}^{(2j)}({\bf B},Z,Z'),
\end{equation}
where $p/p_{2j}$ is set to unity.
This CT potential (a number)
with three arguments is not the operator
${\widehat {\cal U}}_{c(so)}$;
although there is an intimate relationship.
In fact,
\begin{eqnarray}
{\cal U}_{c(so)}^{(2j)}({\bf B},Z,Z') &=&
e^{i(p_{2m}-p)Z} \, e^{-i(p_{2j}-p)Z'}
\nonumber\\* && \qquad \times \sum_{m=0}^\infty
\langle 2m | {\widehat {\cal U}}_{c(so)}({\bf B},Z') |2j \rangle.
\end{eqnarray}

In first-order calculations, a common simple quantity is the integral of
these CT potentials which connect to the nucleon.  Thus,
we define the CT integrals as
\begin{eqnarray}
\Omega_1^{(+)}({\bf B},Z) &=& \int_{-\infty}^Z \frac{dZ'}{2ip}
{\cal U}_{cso}^{(0)}({\bf B},Z,Z'),\\
\Omega_2^{(+)}({\bf B},Z) &=& \frac{B}{2} \int_{-\infty}^Z dZ' \,
{\cal U}_{so}^{(0)}({\bf B},Z,Z'),
\end{eqnarray}
and their derivatives as
\begin{eqnarray}
\Omega_{i,Z}^{(+)}({\bf R})
&=& \frac{\partial}{\partial Z} \Omega_i^{(+)}({\bf R}),\\
\Omega_{i,B}^{(+)}({\bf R})
&=& \frac{\partial}{\partial B} \Omega_i^{(+)}({\bf R}),
\end{eqnarray}
where $i=1,2$.

\subsubsection{Order By Order Wavefunction}

The (outgoing) $OBO$ wavefunction is defined such that
\begin{equation}
\Psi_{CT,{\bf p},-\check{\bf s}_R}^{(+)}({\bf R}) =
e^{ipZ} \sum_m \Psi_{m,{\bf p},-\check{\bf s}_R}^{OBO}({\bf R}),
\end{equation}
where the index $m$ indicates the order (0-2) of the expansion of the path
ordered exponential.
Using standard techniques and previously defined notation, the result is
\begin{equation}
\Psi_{m,{\bf p},-\check{\bf s}_R}^{OBO}
({\bf R}) = {\cal N} \left( \begin{array}{c}
u^{OBO}_{m,{\bf p}}({\bf R}) \\ w^{OBO}_{m,{\bf p}}({\bf R}) \end{array}
\right)
\chi_{-\check{\bf s}_R}.
\end{equation}
where
\begin{eqnarray}
u^{OBO}_{0,{\bf p}} &=&  1
\label{eq:uobo0}\\
w^{OBO}_{0, {\bf p}} &=& \frac{\mbox{\boldmath $\sigma$}
\cdot {\bf p}}{E + M_N},
\label{eq:wobo0}\\
&& \nonumber\\
u^{OBO}_{1,{\bf p}} &=& \left( \Omega_1 + i\Omega_2
\mbox{\boldmath $\sigma$} \cdot \check{\mbox{\boldmath $\phi$}} \right),
\label{eq:uobo1}\\
w^{OBO}_{1,{\bf p}} &=&
\Biggl\{
\frac{\mbox{\boldmath $\sigma$} \cdot {\check{\bf Z}}}{E+M_N} \left[
p \, \Omega_1  +i \frac{\Omega_2}{B}
-i \Omega_{1,Z}+i \Omega_{2,B}
 \right] \nonumber\\*
&&+\, \frac{\mbox{\boldmath $\sigma$}
\cdot {\check{\bf B}}}{E+M_N} \left[ p \, \Omega_2
-i \Omega_{1,B}
-i \Omega_{2,Z}\right] \Biggr\},
\label{eq:wobo1}\\
&&\nonumber\\
u^{OBO}_{2,{\bf p}} &=& \left( {\cal A}_1 + i{\cal A}_2
\mbox{\boldmath $\sigma$} \cdot \check{\mbox{\boldmath $\phi$}} \right),
\label{eq:uobo2}\\
w^{OBO}_{2,{\bf p}} &=&
\Biggl\{
\frac{\mbox{\boldmath $\sigma$} \cdot {\check{\bf Z}}}{E+M_N} \left[
p \, {\cal A}_1  +i \frac{{\cal A}_2}{B}+
-i {\cal A}_{1,Z}
+i {\cal A}_{2,B}
 \right] \nonumber\\*
&&+\, \frac{\mbox{\boldmath $\sigma$}
\cdot {\check{\bf B}}}{E+M_N} \left[ p \, {\cal A}_2
-i {\cal A}_{1,B}
-i {\cal A}_{2,Z}\right] \Biggr\},
\label{eq:wobo2}
\end{eqnarray}
where we have defined two new (outgoing) ``second-order functions'' by
\widetext
\begin{eqnarray}
{\cal A}_1({\bf B},Z) &=& \int_{-\infty}^Z \frac{dZ'}{2ip}
\int_{-\infty}^{Z'} \frac{dZ''}{2ip} \qquad \times
\nonumber \\*
&& \Biggl\{ {\cal U}_{cso}^{(0)}(Z,Z') U_{cso}^{(0,0)}(Z'')
+ B^2 p^2 {\cal U}_{so}^{(0)}(Z,Z') U_{so}^{(0,0)}(Z'') \nonumber\\*
&& + e^{i(p_2-p)(Z'-Z'')} \left(
{\cal U}_{cso}^{(2)}(Z,Z') U_{cso}^{(2,0)}(Z'')
+ B^2 p^2 {\cal U}_{so}^{(2)}(Z,Z') U_{so}^{(2,0)}(Z'') \right) \nonumber \\*
&& + e^{i(p_4-p)(Z'-Z'')} {\cal U}_{cso}^{(4)}(Z,Z') U_{cso}^{(4,0)}(Z'')
\Biggr\},\\
i{\cal A}_2({\bf B},Z) &=& -Bp \int_{\infty}^Z \frac{dZ'}{2ip}
\int_{-\infty}^{Z'} \frac{dZ''}{2ip} \qquad \times
\nonumber \\*
&& \Biggl\{ {\cal U}_{cso}^{(0)}(Z,Z') U_{so}^{(0,0)}(Z'')
+ {\cal U}_{so}^{(0)}(Z,Z') U_{cso}^{(0,0)}(Z'') \nonumber \\*
&& + e^{i(p_2-p)(Z'-Z'')} \left(
{\cal U}_{cso}^{(2)}(Z,Z') U_{so}^{(2,0)}(Z'') +
{\cal U}_{so}^{(2)}(Z,Z') U_{cso}^{(2,0)}(Z'') \right)
\nonumber \\*
&& + e^{i(p_4-p)(Z'-Z'')} {\cal U}_{so}^{(4)}(Z,Z') U_{cso}^{(4,0)}(Z'')
\Biggr\},
\end{eqnarray}
\narrowtext
and we have suppressed the impact parameter inside the arguments of the
CT functions for clarity.

There are a few analytic checks of these expressions.
First,
in the CT limit, when $p_4 \approx p_2 \approx p$ becomes very large, then
$\Omega_{1(2)}$ and ${\cal A}_{1(2)}$
approach zero and CT is obtained.  The next thing to
notice is that
the $OBO_1$
wavefunction has the same form as the $DWBA_1$ with the simple
substitution $\omega \rightarrow \Omega$.  Thus, at low energies (below
excited state threshhold), $p_2$ and $p_4$ are purely imaginary so that
${\cal U}_{c(so)}^{(0)} \rightarrow U_{c(so)}^{(0,0)}$.
Thus, insofar as we can
take $U_{c(so)}^{(0,0)} = U_{c(so)}$, see Eqs.~(\ref{eq:defuc})
and~(\ref{eq:defuso}),
we see that $\Omega \rightarrow
\omega$ and the $DWBA_1$ is obtained, modulo, of course, the small CTD effects
described above.  The same analysis works with the second-order
calculation.  In particular, below threshold, the exponentials damp most of
the terms so that the only ones which contribute are
${\cal U}_{cso(so)}^{(0)} \rightarrow U_{cso(so)}^{(0,0)}$.
Thus, in this low energy limit, ${\cal A}_1 \rightarrow \omega_1^2/2
- \omega_2^2/2$ and ${\cal A}_2 \rightarrow \omega_1 \omega_2$.  Some
simple algebra then yields the desired result that, in the low energy
limit, the $OBO$ wavefunction approaches the $DWBA$.

\subsubsection{Exponential Approximation Wavefunction}

In this approximation, the CT wavefunction is obtained by taking the first
order result and exponentiating.  Thus, the $EA$ has the same relationship
to the $OBO_1$ as does the $DWBA$ to the $DWBA_1$.
This is equivalent to neglecting the path ordering.
Then the (outgoing) $EA$ wavefunction can be written down immediately:
\begin{equation}
\Psi_{CT,{\bf p},-\check{\bf s}_R}^{(+)}
({\bf R}) \approx \Psi^{EA}_{{\bf p},-\check{\bf s}_R}({\bf R}),
\end{equation}
where
\begin{equation}
\Psi^{EA}_{{\bf p},-\check{\bf s}_R}({\bf R}) = {\cal N} \left(
\begin{array}{c}
u^{EA}_{\bf p} ({\bf R}) \\
w^{EA}_{\bf p}({\bf R}) \end{array} \right) \chi_{-\check{\bf s}_R},
\end{equation}
and
\begin{eqnarray}
u^{EA}_{\bf p} &=& e^{ipZ}e^{\Omega_1}
\left[ \cos \Omega_2 + i
\mbox{\boldmath $\sigma$} \cdot \check{\mbox{\boldmath $\phi$}} \sin
\Omega_2 \right],
\label{eq:uea}\\
w^{EA}_{\bf p} &=&
\frac{-i e^{ipZ}e^{\Omega_1}}{E+M_N}
\Biggl\{ \mbox{\boldmath $\sigma$} \cdot
{\check{\bf Z}} \Bigl[ \cos\Omega_2 \left( \Omega_{1,Z}
-\Omega_{2,B}+ip \right)
\nonumber\\* && -\sin\Omega_2 \left(\Omega_{1,B}
+\Omega_{2,Z}-\frac{1}{B} \right) \Bigr] \nonumber\\*
&&+\mbox{\boldmath $\sigma$}
\cdot {\check{\bf B}} \Bigl[ \sin\Omega_2 \left( \Omega_{1,Z}
-\Omega_{2,B}+ip \right)
\nonumber\\* && +\cos\Omega_2 \left(\Omega_{1,B}
+\Omega_{2,Z}\right) \Bigr] \Biggr\}.
\label{eq:wea}
\end{eqnarray}
We again explore the high- and low-energy limits
to perform a check on this wavefunction. At very high energies,
$\Omega_{1(2)}$ become very
small and approach zero.  When this happens, it is easy to see that CT is
obtained.  At very low energies, aside from small quadratic
CTD effects, the $EA$ approaches the $DWBA$.

\subsubsection{Low Energy Wavefunction}

The CT wavefunction in the $LEE$ approximation is defined so that the
expectation value of the CT operators in the nucleon is separated out and
the rest is treated in perturbation theory.  In Ref.~\cite{GM93} we
defined the $LEE$ approximation so that the wavefunction reduced to the
$DWBA$ in the zero'th order.  Due to quadratic and CTD effects, this will
no longer be exactly true.  It is necessary to define more notation to
obtain the
(outgoing) $LEE$ wavefunction.
Thus we define :
\begin{eqnarray}
{\omega_0}_1({\bf B},Z) &\equiv & \int_{-\infty}^Z \frac{dZ'}{2ip}
\biggl[ U_c^{(0,0)}({\bf B},Z')
\nonumber\\* && \qquad \qquad - ipZ' U_{so}^{(0,0)}({\bf B},Z') \biggr],\\
{\omega_0}_2({\bf B},Z) &\equiv & \frac{B}{2} \int_{-\infty}^Z dZ'
U_{so}^{(0,0)}({\bf B},Z'),
\end{eqnarray}
along with the reduced CT integrals
\begin{equation}
{\Omega_0}_{1(2)}({\bf B},Z) =
\Omega_{1(2)}({\bf B},Z) - {\omega_0}_{1(2)}({\bf B},Z).
\end{equation}
With these preliminaries we obtain the (outgoing) $LEE$
wavefunction as
\begin{equation}
\Psi_{CT,{\bf p}, -\check{\bf s}_R}^{(+)}({\bf R}) =
e^{ipZ} \sum_m \Psi_{m,{\bf p},-\check{\bf s}_R}^{LEE}({\bf R}),
\end{equation}
where the index $m$ indicates the order of the expansion of the
perturbation; $m=0$ or 1.
The result is
\begin{equation}
\Psi_{m,{\bf p},-\check{\bf s}_R}^{LEE}
({\bf R}) = {\cal N} \left( \begin{array}{c}
u^{LEE}_{m,{\bf p}}({\bf R}) \\
w^{LEE}_{m,{\bf p}}({\bf R}) \end{array} \right)
\chi_{-\check{\bf s}_R},
\end{equation}
where
\widetext
\begin{eqnarray}
u^{LEE}_{0,{\bf p}} &=& e^{{\omega_0}_1} \left[ \cos {\omega_0}_2 + i
\mbox{\boldmath $\sigma$} \cdot \check{\mbox{\boldmath $\phi$}}
\sin {\omega_0}_2 \right],
\label{eq:ulee0}\\
w^{LEE}_{0,{\bf p}} &=& \frac{e^{ {\omega_0}_1}}{E+M_N} \qquad \times \qquad
\Biggl\{ \nonumber\\*
&& \mbox{\boldmath $\sigma$} \cdot {\check{\bf Z}} \cos {\omega_0}_2
\left( p - i {\omega_0}_{1,Z} +
i {\omega_0}_{2,B} \right)
\nonumber \\* &&
 \mbox{\boldmath $\sigma$} \cdot {\check{\bf Z}} \sin {\omega_0}_2
\left( i {\omega_0}_{1,B} + i{\omega_0}_{2,Z} + \frac{i}{B}\right)
\nonumber\\*
&& \mbox{\boldmath $\sigma$} \cdot {\check{\bf B}} \cos {\omega_0}_2
\left( -i{\omega_0}_{1,B} - i{\omega_0}_{2,Z} \right)
\nonumber\\*
&& \mbox{\boldmath $\sigma$} \cdot {\check{\bf B}} \sin {\omega_0}_2
\left( p - i {\omega_0}_{1,Z} + i {\omega_0}_{2,B} \right)
\Biggr\},
\label{eq:wlee0}\\
\nonumber \\
u^{LEE}_{1,{\bf p}} &=& e^{{\omega_0}_1}
\Bigl[ \left(\cos {\omega_0}_2 + i
\mbox{\boldmath $\sigma$} \cdot \check{\mbox{\boldmath $\phi$}}
\sin {\omega_0}_2 \right) {\Omega_0}_1  + \nonumber \\*
&& \qquad \qquad \left( -\sin {\omega_0}_2
+ i \mbox{\boldmath $\sigma$} \cdot \check{\mbox{\boldmath $\phi$}}
\cos {\omega_0}_2 \right) {\Omega_0}_2
\Bigr]
\label{eq:ulee1},\\
w^{LEE}_{1,{\bf p}}
&=& \frac{e^{{\omega_0}_1}}{E+M_N} \qquad \times \qquad
\Biggl\{
\nonumber\\*
&& \mbox{\boldmath $\sigma$} \cdot {\check{\bf Z}} \cos {\omega_0}_2
\bigl[ \left( p - i {\omega_0}_{1,Z} +
i {\omega_0}_{2,B} \right)
{\Omega_0}_1 +
\nonumber \\* &&
i \left( {\omega_0}_{1,B} + {\omega_0}_{2,Z} + \frac{1}{B}
\right) {\Omega_0}_2
- i {\Omega_0}_{1,Z}
+ i {\Omega_0}_{2,B}\bigr]\nonumber\\*
&& \mbox{\boldmath $\sigma$} \cdot {\check{\bf Z}} \sin {\omega_0}_2
\bigl[ \left( -p + i {\omega_0}_{1,Z} - i {\omega_0}_{2,B} \right)
{\Omega_0}_2 + \nonumber\\*
&&
i \left( {\omega_0}_{1,B} + {\omega_0}_{2,Z} + \frac{1}{B}
\right)  {\Omega_0}_1
+ i {\Omega_0}_{1,B}
+ i{\Omega_0}_{2,Z} \bigr]
\nonumber\\*
&& \mbox{\boldmath $\sigma$} \cdot {\check{\bf B}} \cos {\omega_0}_2
\bigl[ \left( p - i {\omega_0}_{1,Z} + i {\omega_0}_{2,B} \right)
{\Omega_0}_2 - \nonumber \\*
&&
i \left( {\omega_0}_{1,B} + {\omega_0}_{2,Z} \right) {\Omega_0}_1
- i {\Omega_0}_{1,B}
- i{\Omega_0}_{2,Z} \bigr]
\nonumber\\*
&& \mbox{\boldmath $\sigma$} \cdot {\check{\bf B}} \sin {\omega_0}_2
\bigl[ \left( p - i {\omega_0}_{1,Z} + i {\omega_0}_{2,B} \right)
{\Omega_0}_1 + \nonumber \\*
&&
i \left( {\omega_0}_{1,B}
+ {\omega_0}_{2,Z}
\right) {\Omega_0}_2
- i {\Omega_0}_{1,Z}
+ i {\Omega_0}_{2,B}\bigr] \Biggr\},
\label{eq:wlee1}\\
\nonumber\\
u^{LEE}_{2,{\bf p}} &=& e^{{\omega_0}_1}
\Bigl[ \left(\cos {\omega_0}_2 + i
\mbox{\boldmath $\sigma$} \cdot \check{\mbox{\boldmath $\phi$}}
\sin {\omega_0}_2 \right) {\cal B}_1  + \nonumber \\*
&& \qquad \quad \left( -\sin {\omega_0}_2
+ i \mbox{\boldmath $\sigma$} \cdot \check{\mbox{\boldmath $\phi$}}
 \cos {\omega_0}_2 \right) {\cal B}_2
\Bigr]
\label{eq:ulee2},\\
w^{LEE}_{2,{\bf p}}
&=& \frac{e^{{\omega_0}_1}}{E+M_N} \qquad \times \qquad
\Biggl\{
\nonumber\\*
&& \mbox{\boldmath $\sigma$} \cdot {\check{\bf Z}} \cos {\omega_0}_2
\bigl[ \left( p - i {\omega_0}_{1,Z} +
i {\omega_0}_{2,B} \right)
{\cal B}_1 +
\nonumber \\* &&
i \left( {\omega_0}_{1,B} + {\omega_0}_{2,Z} + \frac{1}{B}
\right) {\cal B}_2
- i {\cal B}_{1,Z}
+ i {\cal B}_{2,B} \bigr]\nonumber\\*
&& \mbox{\boldmath $\sigma$} \cdot {\check{\bf Z}} \sin {\omega_0}_2
\bigl[ \left( -p + i {\omega_0}_{1,Z} - i {\omega_0}_{2,B} \right)
{\cal B}_2 + \nonumber\\*
&&
i \left( {\omega_0}_{1,B} + {\omega_0}_{2,Z} + \frac{1}{B}
\right)  {\cal B}_1
+ i {\cal B}_{1,B}
+ i{\cal B}_{2,Z} \bigr]
\nonumber\\*
&& \mbox{\boldmath $\sigma$} \cdot {\check{\bf B}} \cos {\omega_0}_2
\bigl[ \left( p - i {\omega_0}_{1,Z} + i {\omega_0}_{2,B} \right)
{\cal B}_2 - \nonumber \\*
&&
i \left( {\omega_0}_{1,B} + {\omega_0}_{2,Z} \right) {\cal B}_1
- i {\cal B}_{1,B}
- i{\cal B}_{2,Z} \bigr]
\nonumber\\*
&& \mbox{\boldmath $\sigma$} \cdot {\check{\bf B}} \sin {\omega_0}_2
\bigl[ \left( p - i {\omega_0}_{1,Z} + i {\omega_0}_{2,B} \right)
{\cal B}_1 + \nonumber \\*
&&
i \left( {\omega_0}_{1,B}
+ {\omega_0}_{2,Z}
\right) {\cal B}_2
- i {\cal B}_{1,Z}
+ i {\cal B}_{2,B} \bigr] \Biggr\},
\label{eq:wlee2}
\end{eqnarray}
where we have defined two new functions, in analogy to ${\cal A}_{1(2)}$,
to be
\widetext
\begin{eqnarray}
{\cal B}_1({\bf B},Z) &=& \int_{-\infty}^Z \frac{dZ'}{2ip}
\int_{-\infty}^{Z'} \frac{dZ''}{2ip}
\Biggl\{ e^{i(p_2-p)(Z'-Z'')} \, ^{(0,0)}{\cal U}_{cso}^{(2)}(Z,Z')
U_{cso}^{(2,0)}(Z'') \nonumber \\*
&& + B^2 p^2 e^{i(p_2-p)(Z'-Z'')} \, ^{(0,0)}{\cal U}_{so}^{(2)}(Z,Z')
U_{so}^{(2,0)}(Z'') \nonumber \\*
&& + e^{i(p_4-p)(Z'-Z'')} \, ^{(0,0)}{\cal U}_{cso}^{(4)}(Z,Z')
U_{cso}^{(4,0)}(Z'') \Biggr\},
\label{eq:lee2fun1}\\
i {\cal B}_2({\bf B},Z) &=& -Bp \int_{-\infty}^Z \frac{dZ'}{2ip}
\int_{-\infty}^{Z'} \frac{dZ''}{2ip}
\Biggl\{ e^{i(p_2-p)(Z'-Z'')} \, ^{(0,0)}{\cal U}_{so}^{(2)}(Z,Z')
U_{cso}^{(2,0)}(Z'') \nonumber \\*
&& + e^{i(p_2-p)(Z'-Z'')} \, ^{(0,0)}{\cal U}_{cso}^{(2)}(Z,Z')
U_{so}^{(2,0)}(Z'') \nonumber \\*
&& + e^{i(p_4-p)(Z'-Z'')} \, ^{(0,0)}{\cal U}_{so}^{(4)}(Z,Z')
U_{cso}^{(4,0)}(Z'') \Biggr\}.
\end{eqnarray}
\narrowtext
We have also defined the new functions
\begin{eqnarray}
^{(0,0)}{\cal U}_{cso(so)}^{(2j)}(Z,Z') &=& {\cal U}_{cso(so)}(Z,Z')
\nonumber\\* &&  -
e^{i(p_{2j}-p)(Z-Z')} U_{cso(so)}^{(0,0)}(Z').
\end{eqnarray}

The behavior of this
$LEE$ wavefunction at low energies has already been
briefly discussed.  Note that at very low energies,
${\Omega_0}_i \rightarrow 0$ as well as
$^{(0,0)}{\cal U}_{cso(so)} \rightarrow 0$.
Therefore, except for quadratic and CTD effects,
the $LEE$ wavefunction approaches the $DWBA$ wavefunction for low
energies.
One can also check this
wavefunction at high energies where
the potentials ${\Omega_0}_i
\rightarrow -{\omega_0}_i$.  Now, this wavefunction claims
to be correct only to second order in the CT potentials.
Thus, if we
make the replacement that ${\Omega_0}_i
\rightarrow -{\omega_0}_i$, we can expand the
resulting wavefunction to first order in the potentials ${\omega_0}_i$.
Then, in addition,
letting $\cos {\omega_0}_2
\rightarrow 1 - {\omega_0}_2^2/2$,
$\sin {\omega_0}_2 \rightarrow {\omega_0}_2$ and
$e^{{\omega_0}_1} \rightarrow 1+{\omega_0}_1 + {\omega_0}_1^2/2$
it is relatively simple to show that,
to within terms of ${\cal O}({\omega_0}^3)$,
the $LEE_1$ wavefunction approaches
the plane wave result for very large momenta.  In case the reader wishes to
verify the above statement, an intermediate result is that, in this limit,
the ${\cal B}_1 = \left( {\omega_0}_1^2 - {\omega_0}_2^2 \right)/2
+ {\cal O}(\omega_0^3)$ and ${\cal B}_2 = {\omega_0}_1 \, {\omega_0}_2
+ {\cal O}(\omega_0^3)$.

\begin{figure}
\caption{Schematic drawing of $(\vec e, e' \vec p)$ reaction.}
\label{fig:veep}
\end{figure}

\begin{figure}
\caption{Coordinate system used to describe the
$(\vec e,e' \vec N)$ reaction.}
\label{fig:kinematics}
\end{figure}

\begin{figure}
\caption{The second-order cross section ratios; $M_2=1.44 \, GeV$.}
\label{fig:o2l}
\end{figure}

\begin{figure}
\caption{The second-order cross section ratios; $M_2=1.80 \, GeV$.}
\label{fig:o2h}
\end{figure}

\begin{figure}
\caption{Total cross section ratios for $^{12}C$.  Solid (circles) is DWBA,
dotdashed (diamonds) includes CT, via EA, for $M_2=1.44 \, GeV$, dashed
(boxes) have $M_2=1.80 \, GeV$.}
\label{fig:c12tcs}
\end{figure}

\begin{figure}
\caption{Total cross section ratios for $^{40}Ca$.
The curves are as in Figure~\protect\ref{fig:c12tcs}. }
\label{fig:ca40tcs}
\end{figure}

\begin{figure}
\caption{Total cross section ratios for $^{208}Pb$.
The curves are as in Figure~\protect\ref{fig:c12tcs}. }
\label{fig:pb208tcs}
\end{figure}

\begin{figure}
\caption{Ratio of integrated longitudinal to transverse response for
$^{12}C$.  Dotted (fancy boxes) is plane wave result, solid (circles) is
DWBA, dotdashed (diamonds) and dashes (boxes) are CT-included cases for
light and heavy excited state masses.}
\label{fig:c12rlrt}
\end{figure}

\begin{figure}
\caption{Ratio of integrated longitudinal to transverse response for
$^{40}Ca$.  The curves are as in Figure~\protect\ref{fig:c12rlrt}. }
\label{fig:ca40rlrt}
\end{figure}

\begin{figure}
\caption{Ratio of integrated longitudinal to transverse response for
$^{208}Pb$.  The curves are as in Figure~\protect\ref{fig:c12rlrt}. }
\label{fig:pb208rlrt}
\end{figure}

\begin{figure}
\caption{Current conservation violations for $^{12}C$.  The
curves are as in Figure~\protect\ref{fig:c12rlrt}. }
\label{fig:c12drl}
\end{figure}

\begin{figure}
\caption{Current conservation violations for $^{40}Ca$.  The
curves are as in Figure~\protect\ref{fig:c12rlrt}. }
\label{fig:ca40drl}
\end{figure}

\begin{figure}
\caption{Current conservation violations for $^{208}Pb$.  The
curves are as in Figure~\protect\ref{fig:c12rlrt}. }
\label{fig:pb208drl}
\end{figure}

\begin{figure}
\caption{Differential unpolarized
cross section and normal polarization for $^{12}C$ at
${Q^2=0.96 \, GeV^2}$.  The curves are as in
Figure~\protect\ref{fig:c12rlrt}.}
\label{fig:c12s0pn1}
\end{figure}

\begin{figure}
\caption{Differential unpolarized
cross section and normal polarization for $^{12}C$ at
${Q^2=20.86 \, GeV^2}$.  The curves are as in
Figure~\protect\ref{fig:c12rlrt}.}
\label{fig:c12s0pn20}
\end{figure}

\begin{figure}
\caption{Effect of different
electron kinematics at $Q^2 = 5.96 \, GeV^2$.
The dotdashed curve is the $EA$ for $M_2=1.44\, GeV$, the dashed curve is
the $OBO_1$, the dotted curve is the $LEE_1$.  The solid curves are the
$DWBA$ and the plane wave result, the context identifies them.}
\label{fig:other6}
\end{figure}

\begin{figure}
\caption{Differential unpolarized response functions for $^{208}Pb$ at
$Q^2 = 3.25 \, GeV^2$.  The curves are as
in Figure~\protect\ref{fig:c12rlrt}.}
\label{fig:pb208r3}
\end{figure}

\begin{figure}
\caption{Differential normal response functions for $^{208}Pb$ at
$Q^2 = 3.25 \, GeV^2$.  The curves are as
in Figure~\protect\ref{fig:c12rlrt}.}
\label{fig:pb208rn3}
\end{figure}

\begin{figure}
\caption{Cross section ratios for finite-sized
wavepacket.}
\label{fig:c12tcsfs}
\end{figure}

\begin{figure}
\caption{Current conservation violations for
finite-sized wavepacket.}
\label{fig:c12drlfs}
\end{figure}

\begin{figure}
\caption{Effect of Fermi motion.}
\label{fig:kz}
\end{figure}

\begin{figure}
\caption{Total cross
section ratio for $^{12}C$ with no lower components.  The upper CT
curve and the lower DWBA curve are calculated with wavefunctions having no
lower components.  The point is that the ratio is independent of the lower
components.}
\label{fig:c12tcslc}
\end{figure}

\begin{figure}
\caption{Current conservation violations for $^{12}C$ with no lower
components.
In this figure, the dotted curve is the Born calculation with no lower
components; the other curves are as in Figure~\protect\ref{fig:c12tcslc}.}
\label{fig:c12drllc}
\end{figure}

\begin{figure}
\caption{Ratio of integrated longitudinal to
transverse response for $^{12}C$ with no lower components.
The curves are as in Figure~\protect\ref{fig:c12tcslc}.}
\label{fig:c12rlrtlc}
\end{figure}

\begin{table}
\caption{\label{tab:strengths} Energy dependent strengths of optical
potentials.  $Q^2$ is in $GeV^2$ and $p_{lab}$ is in $GeV$.
$r=-4\pi i p_{lab}/M_N$.}
\begin{tabular}{
r@{}l
r@{}l
r@{}l@{$+$}r@{}l
r@{}l@{$-$}r@{}l
r@{}l@{$+$}r@{}l
r@{}l@{$-$}r@{}l
}
\multicolumn{2}{c}{$Q^2$}&
\multicolumn{2}{c}{$p_{lab}$} &
\multicolumn{4}{c}{$r\,F_s^0 (fm^2)$} &
\multicolumn{4}{c}{$r \, F_v^0 (fm^2)$} &
\multicolumn{4}{c}{$\rho_0 \, r \,F_s^0 (MeV)$} &
\multicolumn{4}{c}{$\rho_0 \, r \, F_v^0 (MeV)$} \\
\tableline
$0$ & $.96$ &
$1$ & $.1$  &
$  -9$ & $.566$ & $ 1$ & $.768 \, i$ &
$   5$ & $.720$ & $ 2$ & $.291 \, i$ &
$-313$ & $.3  $ & $57$ & $.91  \, i$ &
$ 187$ & $.4  $ & $75$ & $.05  \, i$ \\
$1$ & $.88$ &
$1$ & $.7$ &
$  -9$ & $.550$ & $  3$ & $.748\, i$ &
$   4$ & $.970$ & $  3$ & $.895\, i$ &
$-312$ & $.8  $ & $122$ & $ .8 \, i$ &
$ 162$ & $.8  $ & $127$ & $ .6 \, i$ \\
$2$ & $.38$ &
$2$ & $.0$ &
$-10$ & $.46$ & $4$ & $.219 \, i$ &
$4$ & $.912$ & $3$ & $.936 \,i$ &
$-342$ & $.6$ & $138$ & $.2 \, i$ &
$160$ & $.9$ & $128$ & $.9 \, i$ \\
$3$ & $.25$ &
$2$ & $.5$ &
$-7$ & $.243$ & $4$ & $.047 \, i$ &
$3$ & $.396$ & $3$ & $.573 \, i$ &
$-237$ & $.3$ & $132$ & $.6 \, i$ &
$111$ & $.2$ & $117$ & $.0 \, i$ \\
$4$ & $.14$ &
$3$ & $.0$ &
$-10$ & $.75$ & $6$ & $.735 \, i$ &
$3$ & $.836$ & $4$ & $.124 \, i$ &
$-352$ & $.0$ & $220$ & $.6 \, i$ &
$125$ & $.6$ & $135$ & $.1 \, i$ \\
$5$ & $.96$ &
$4$ & $.0$ &
$-7$ & $.091$ & $2$ & $.147 \, i$ &
$2$ & $.155$ & $2$ & $.540 \, i$ &
$-232$ & $.3$ & $70$ & $.33 \, i$ &
$70$ & $.59$ & $83$ & $.20 \, i$ \\
$9$ & $.65$ &
$6$ & $.0$ &
$-7$ & $.681$ & $2$ & $.398 \, i$ &
$1$ & $.831$ & $2$ & $.380 \, i$ &
$-251$ & $.6$ & $78$ & $.56 \, i$ &
$59$ & $.98$ & $77$ & $.96 \, i$ \\
$20$ & $.86$ &
$12$ & $.0$ &
$-7$ & $.631$ & $10$ & $.73 \, i$ &
$1$ & $.168$ & $2$ & $.809 \, i$ &
$-250$ & $.0$ & $351$ & $.4 \, i$ &
$38$ & $.27$ & $92$ & $.03 \, i$\\
\end{tabular}
\end{table}

\begin{table}
\caption{\label{tab:csr}
Cross section and ratio of real to imaginary parts of forward
scattering amplitude, as given by optical potential strengths.}
\begin{tabular}{r@{}lr@{}lr@{}lr@{}l}
\multicolumn{2}{c}{$Q^2$}&
\multicolumn{2}{c}{$E_{lab}$} &
\multicolumn{2}{c}{$\sigma$} &
\multicolumn{2}{c}{$\alpha_f$} \\
\multicolumn{2}{c}{$(GeV^2)$}&
\multicolumn{2}{c}{$(GeV)$} &
\multicolumn{2}{c}{$(mb)$} &
\multicolumn{2}{c}{(unitless)} \\ \tableline
$0$ & $.96$ & $1$ & $.45$ & $30$ & $.1$ & $0$ & $.43$ \\
$1$ & $.88$ & $1$ & $.94$ & $47$ & $.7$ & $-0$ & $.17$ \\
$2$ & $.38$ & $2$ & $.21$ & $47$ & $.5$ & $-0$ & $.22$ \\
$3$ & $.25$ & $2$ & $.67$ & $46$ & $.0$ & $-0$ & $.40$ \\
$4$ & $.14$ & $3$ & $.14$ & $44$ & $.3$ & $-0$ & $.30$ \\
$5$ & $.96$ & $4$ & $.11$ & $42$ & $.2$ & $-0$ & $.26$ \\
$9$ & $.65$ & $6$ & $.07$ & $40$ & $.8$ & $-0$ & $.32$ \\
$20$ & $.86$ & $12$ & $.04$ & $39$ & $.6$ & $-0$ & $.29$ \\
\end{tabular}
\end{table}

\begin{table}
\caption{\label{tab:kin}Calculational kinematics.}
\begin{tabular}{r@{}lr@{}lr@{}lr@{}lr@{}lr@{}lr@{}lr@{}l}
\multicolumn{2}{c}{$Q^2$} &
\multicolumn{2}{c}{$E_i$} &
\multicolumn{2}{c}{$\theta_e$} &
\multicolumn{2}{c}{$\left( \frac{d\sigma}{d\Omega_{k'}}\right)_{Mott}$} &
\multicolumn{2}{c}{$V_L$} &
\multicolumn{2}{c}{$V_T$} &
\multicolumn{2}{c}{$V_{TT}$} &
\multicolumn{2}{c}{$V_{LT}$} \\
\multicolumn{2}{c}{$(GeV^2)$} &
\multicolumn{2}{c}{$(GeV)$} &
\multicolumn{2}{c}{(deg)} &
\multicolumn{2}{c}{(nb)} & & & & \\ \tableline
0&.96&4&.0&15&.1&1091&.&0&.62&0&.41&0&.39&$-$0&.70\\
1&.88&4&.0&22&.8&203&.0&0&.43&0&.37&0&.33&$-$0&.54\\
2&.38&4&.0&27&.0&103&.6&0&.36&0&.36&0&.30&$-$0&.48\\
3&.25&4&.0&34&.8&36&.90&0&.27&0&.36&0&.26&$-$0&.41\\
4&.14&4&.0&44&.6&13&.40&0&.21&0&.40&0&.23&$-$0&.36\\
5&.96&4&.0&84&.5&0&.8865&0&.14&1&.01&0&.19&$-$0&.41\\
5&.96&6&.0&34&.5&17&.07&0&.14&0&.28&0&.19&$-$0&.25\\
5&.96&11&.0&22&.7&140&.7&0&.14&0&.20&0&.19&$-$0&.24\\
9&.65&15&.0&14&.7&85&.35&0&.072&0&.15&0&.13&$-$0&.14\\
20&.86&21&.0&18&.2&18&.24&0&.021&0&.098&0&.072&$-$0&.060\\
\end{tabular}
\end{table}

\begin{table}
\caption{\label{tab:fsdata}
Forward scattering data at $p_{lab}=6$ and $12 \, GeV/c$.}
\begin{tabular}{llr@{}lcllr@{}l}
\multicolumn{9}{c}{$p_{lab}=6\, GeV/c$} \\ \tableline
$\sigma^{tot}$ & = & 40 &.75 $mb$ &  & $\alpha_f$ & = & $-0$&.32\\
$\Delta \sigma_T$ &=&0&.35 $mb$ &  &$F_2$ &=& $-$4&.60 $GeV^{-1}$ \\
$\Delta \sigma_L$ &=& $-1$&.04 $\, mb$  && $F_3$ &=& 4&.60 $GeV^{-1}$ \\
$\left( \frac{d\sigma}{dt} \right)_{t=0}$ &=& 93&.0 $mb$ &&
$\left( \frac{dP_n}{dq} \right)_{q=0}$ &=& 0&.494 $GeV^{-1}$ \\ \tableline
\multicolumn{9}{c}{$p_{lab}=12\, GeV/c$} \\ \tableline
$\sigma^{tot}$ & = & 39&.60 $mb$ &  & $\alpha_f$ & = & $-0$&.29 \\
$\Delta \sigma_T$ &=&0&.01 $mb$ &  &$F_2$ &=& $-$6&.05 $GeV^{-1}$ \\
 $\Delta \sigma_L$ &=& $-0$&.73\, $mb$ && $F_3$ &=& 3&.55 $GeV^{-1}$ \\
$\left( \frac{d\sigma}{dt} \right)_{t=0}$ &=& 65&.0 $mb$ &&
$\left( \frac{dP_n}{dq} \right)_{q=0}$ &=& 0&.215 $GeV^{-1}$\\
\end{tabular}
\end{table}

\begin{table}
\caption{\label{tab:pauliamps} Pauli amplitudes for $pp$ scattering}
\begin{tabular}{rcccc}
\multicolumn{1}{c}{$Q^2$} & $A^0$ & $B^0$ & $C^0$ & $E^0$ \\
\multicolumn{1}{c}{$(GeV^2)$}
& $(GeV^{-2})$ & $(GeV^{-2})$ & $(GeV^{-3})$ & $(GeV^{-2})$
\\ \tableline
$0.96$&$3.08-1.31\, i$&$-0.22+1.21\, i$&$-1.76-6.55\, i$&$0.69+0.02\, i$\\
$1.88$&$4.87+0.81\, i$&$-0.23+1.03\, i$&$-3.13-5.44\, i$&$0.82-0.05\, i$\\
$2.38$&$4.85+1.06\, i$&$-0.38+0.71\, i$&$-3.11-5.22\, i$&$0.81+0.16\, i$\\
$3.25$&$4.70+1.86\, i$&$-0.14+0.48\, i$&$-2.81-3.35\, i$&$0.29+0.16\, i$\\
$4.14$&$4.53+1.34\, i$&$-0.05+0.47\, i$&$-3.41-3.95\, i$&$0.15-0.01\, i$\\
$5.96$&$4.31+1.13\, i$&$-0.04+0.40\, i$&$-1.62-2.13\, i$&$0.12-0.09\, i$\\
$9.65$&$4.16+1.33\, i$&$-0.02+0.19\, i$&$-1.34-1.62\, i$&$0.07-0.00\, i$\\
$20.86$&$4.05+1.17\, i$&$-0.00+0.13\, i$&$-1.61-0.87\, i$&$0.04-0.05\, i$\\
\end{tabular}
\end{table}

\end{document}